\documentclass[prd,10pt,aps,twocolumn,superscriptaddress,floatfix,notitlepage,nofootinbib,amssymb,amsmath]{revtex4-1}
\usepackage{slashed}

\usepackage{ulem}
\renewcommand\sout{\bgroup \color{red} \ULdepth=-.5ex \ULset}

\usepackage{mathtools}
\usepackage{epsfig}
\usepackage{bm}

\newcommand{\tE}{{\widetilde E}}

\newcommand{\ed}{{\mathrm {edge}}}
\newcommand{\bulk}{{\mathrm{bulk}}}
\newcommand{\mM}{{\mathcal{M}}}
\newcommand{\sign}[1]{{\mathrm{sign}}(#1)}
\newcommand{\beqn}{\begin{eqnarray}}
\newcommand{\eeqn}{\end{eqnarray}}
\newcommand{\eq}[1]{(\ref{#1})}
\newcommand{\jj}{{\mathrm{j}}}

\newcommand{\bb}{\mathsf{b}}
\newcommand{\bp}{\mathsf{p}}
\newcommand{\bE}{\mathsf{E}}

\newcommand{\lab}{{\mathrm {lab}}}

\newcommand{\Z}{{\mathbb Z}}
\newcommand{\R}{{\mathbb R}}

\newcommand{\dirac}{\slashed \partial}
\newcommand{\bs}{\boldsymbol}

\newcommand{\avr}[1]{{\left\langle #1 \right\rangle}}

\def\bbbone{{\mathchoice {\rm 1\mskip-4mu l} {\rm 1\mskip-4mu l} {\rm 1\mskip-4.5mu l} {\rm 1\mskip-5mu l}}}

\begin{document}

\title{Edge states and thermodynamics of rotating relativistic fermions under magnetic field}

\author{M. N. Chernodub}
\affiliation{Laboratoire de Math\'ematiques et Physique Th\'eorique UMR 7350, Universit\'e de Tours, Tours 37200 France}
\affiliation{Laboratory of Physics of Living Matter, Far Eastern Federal University, Sukhanova 8, Vladivostok, 690950, Russia}
\author{Shinya Gongyo}
\affiliation{Theoretical Research Division, Nishina Center, RIKEN, Wako, Saitama 351-0198, Japan}

\date{\today}

\begin{abstract}
We discuss free Dirac fermions rotating uniformly inside a cylindrical cavity in the presence of background magnetic field parallel to the cylinder axis. We show that in addition to the known bulk states the system contains massive edge states with the masses inversely proportional to the radius of the cylinder. The edge states appear at quantized threshold values of the fermion mass. In the limit of infinite fermion mass the masses of the edge states remain finite but, generally, nonzero as contrasted to the bulk states whose masses become infinite. The presence of magnetic field affects the spectrum of both bulk and edge modes, and the masses of the edge states may vanish at certain values of magnetic field. The moment of inertia of Dirac fermions is non-monotonically increasing, oscillating function of magnetic field. The oscillations are well pronounced in a low-temperature domain and they disappear at high temperatures.
\end{abstract}

\maketitle

\section{Introduction}

Rotating systems of relativistic fermions appear in various physical settings characterized by different energy scales. The examples include interior of rapidly spinning neutron stars~\cite{Cook:1993qr}, quark-gluon plasma in noncentral heavy-ion collisions~\cite{ref:HIC}, and anomalous chiral transport phenomena~\cite{ref:CVE} applied both to neutrino fluxes in rotating astrophysical environments~\cite{Vilenkin:1980zv,ref:Vilenkin} and to semimetal materials in solid state applications~\cite{ref:Weyl}. 

Rotation changes the spectrum of free fermions~\cite{Iyer:1982ah,ref:Becattini,Ambrus:2014uqa,Ambrus:2015lfr,Manning:2015sky} and, consequently, affects the mass gap generation in interacting fermionic systems. For example, the critical temperature of chiral symmetry restoration $T_c$ is a diminishing function of the rotational angular frequency~$\Omega$ \cite{ref:McInnes,Chen:2015hfc,Jiang:2016wvv,Ebihara:2016fwa,Chernodub:2016kxh}. The rotational effects have been studied under simplifying assumption that the rotation is globally uniform, so that the angular velocity does not depend on the distance to the rotational axis. A uniformly rotating relativistic system should be bounded in the transverse directions with respect to the axis of rotation in order to comply with the causality principle. The latter requires that the velocity of particles should not exceed the speed of light to avoid pathological effects~\cite{Ambrus:2014uqa,ref:superluminal}. The presence of the boundary implies a dependence of the chiral restoration temperature $T_c = T_c(\Omega)$ on geometrical features, in particular, on the type of the boundary condition~\cite{Chernodub:2017ref}. The uniform rotation in magnetic field background but in an unrestricted transverse geometry has been studied in Ref.~\cite{Chen:2015hfc}.

In this paper we generalize the results of Refs.~\cite{Ambrus:2015lfr,Chernodub:2017ref} in threefold way. Firstly, we show that in addition to the bulk modes the spectrum of free massive Dirac fermions contains the edge states localized at the boundary of the cylinder. Secondly, we discuss the spectrum of both bulk and edge modes in the presence of external magnetic field. Finally, we illustrate the importance of the edge modes for thermodynamics of free Dirac fermions and for its rotational properties such as moment of inertia which exhibits curious oscillating behavior as a function of magnetic field.

Notice that possible effects of the edge states were not accounted for in existing studies of phase structure of the interacting rotating fermions~\cite{Chen:2015hfc,Ebihara:2016fwa,Jiang:2016wvv,Chernodub:2016kxh,Chernodub:2017ref}. In Ref.~\cite{Chen:2015hfc,Jiang:2016wvv} rotational properties were investigated in the transversally unrestricted geometry which questions the consistency with the requirement of relativistic causality under uniform rotation and, simultaneously, does not allow for the presence of the edge states. The existence of the edge states, found in the present paper, definitely calls for a re-estimation of the phase diagram of interacting fermions under uniform rotation. 

We would like to mention that in solid state terms the system of Dirac fermions considered in this article corresponds to a non-topological insulator as it is characterized by the presence of gapped bulk modes and the absence of symmetry-protected boundary (edge) states with zero mass. The edge states are generally massive and their mass is proportional to the mean curvature of the cylinder surface. This statement is not surprising because the Dirac equation alone is not enough to describe topological insulators~\cite{ref:Shen}, where the presence of zero-mass edge states is guaranteed by topological reasons of underlying lattice Hamiltonians~\cite{ref:TI}.

The structure of this paper is as follows. In Sect.~\ref{sec:rotating} we review, following Ref.~\cite{Ambrus:2015lfr}, known bulk solutions for the Dirac fermions in the cylinder with the MIT boundary conditions in the absence of magnetic field. We also discuss particularities of the spectrum for the chiral boundary conditions~\cite{Chernodub:2017ref}. In the same section we find the edge states of the system and describe their properties. In Sect.~\ref{sec:rotating:magnetic} we discuss properties of bulk and edge solutions in the magnetic field background. Section~\ref{sec:edge:rotation} is devoted to studies of rotational properties of the system in the limit of (negative) infinite fermion mass. In this limit the thermodynamics of the system is given by the edge states only, allowing us to highlight the importance of the edge states. The last section is devoted to conclusions. 

\clearpage

\section{Bulk and edge solutions in the absence of magnetic field}
\label{sec:rotating}

In this section we discuss solutions of massive rigidly rotating Dirac fermions confined in a cylindrical geometry in the absence of magnetic field. We start from the known bulk states that were already described in Ref.~\cite{Ambrus:2015lfr}  (see also Ref.~\cite{Chernodub:2016kxh}) and then we demonstrate that the system contains also certain new (edge) states which possess rather peculiar properties.

\subsection{Dirac equation in the cylinder}

We consider a system of free fermions which is rigidly rotating with the angular frequency $\Omega$ about the axis of the infinitely long cylinder of the radius $R$. 

Given the geometry of the system it is convenient to work in the cylindrical coordinates, $x \equiv (x_0, x_1,x_2,x_3) = (t,\rho\cos\varphi,\rho\sin\varphi,z)$. There are two natural reference frames in this problem: the inertial laboratory  frame and non-inertial corotating  frame. The former one corresponds to a rest frame while the latter one is rigidly fixed with the rotating system. The coordinates $t$, $\rho$ and $z$ in the corotating reference frame coincide with the corresponding coordinates of the laboratory frame: $t = t_{\lab}$, $\rho = \rho_\lab$ and $z = z_\lab$. The angular variables in these frames are related as follows: 
\beqn
\varphi = [\varphi_{\mathrm{lab}} - \Omega t]_{2\pi}\,,
\label{eq:varphi}
\eeqn
where $[\dots]_{2\pi}$ means ``modulo $2\pi$''. The simple relation between angular variables~\eq{eq:varphi} leads, nevertheless, to quite nontrivial metric in the corotating frame:
\begin{align}
g_{\mu \nu} =
\left(
\begin{array}{cccc}
1-(x^2+y^2)\Omega ^2 & y\Omega & -x\Omega & 0 \\
y\Omega & -1 & 0 & 0 \\
-x\Omega & 0 & -1 & 0 \\
0 & 0 & 0 &-1
\end{array}
\right),
\label{eq:g:mu:nu:matrix}
\end{align}
which corresponds to the line element
\beqn
ds^2 & = & g_{\mu\nu} dx^\mu dx^\nu = \eta_{\hat\mu\hat\nu} dx^{\hat \mu} dx^{\hat \nu}
\label{eq:metric} \\
& = & \left(1-\rho ^2 \Omega ^2 \right)dt^2- 2\rho^2\Omega dt d\varphi - d\rho ^2- \rho^2 d\varphi^2 - d z^2\,,
\nonumber
\eeqn
where $\eta_{\hat\mu\hat\nu} = {\mathrm{diag}}\, (1,-1,-1,-1)$ is the flat metric. Here we adopt the convention that $\hat{i},\hat{j} \dots = \hat{t},\hat{x},\hat{y},\hat{z}$ and $\mu,\nu \dots = t, x, y, z$ refer to the local coordinates in the laboratory frame and the corotating frame, respectively. We use the units in which the speed of light and the reduced Planck constant are equal to unity, $c = \hbar = 1$.

The spectrum of the fermions is described by the eigenfunctions of the free Dirac equation in the corotating reference frame:
\beqn
\left[ i \gamma^\mu \left(\partial_\mu + \Gamma^\mu \right)- M \right] \psi = 0\,,
\label{eq:Dirac:rotating}
\eeqn
where the Dirac matrices in the curved corotating space-time $\gamma^\mu(x) = e^{\mu} _{\hat{i}}(x) \gamma ^{\hat{i}}$ are connected to the matrices in the laboratory frame $\gamma ^{\hat{i}}$ via the vierbein $e^{\mu}_{\hat{i}}$. The vierbein is a ``square root'' of the metric $\eta_{\hat{i} \hat{j}} = g_{\mu \nu}e^{\mu}_{\hat{i}}e^{\nu}_{\hat{j}}$. In the case of metric~\eq{eq:g:mu:nu:matrix} the vierbein may be chosen in the form 
\beqn
e^t_{\hat{t}}=e^x_{\hat{x}}=e^y_{\hat{y}}=e^y_{\hat{y}}=1,
\quad 
e^x_{\hat{t}}= y\Omega, 
\quad
e^y_{\hat{t}}= -x\Omega, \quad
\eeqn
with all other components of $\eta_{\hat{i} \hat{j}}$ being zero.

In Eq.~\eq{eq:Dirac:rotating} the spin connection $\Gamma^\mu$ in the metric~\eq{eq:g:mu:nu:matrix} has only one nonzero component:
\begin{align}
\Gamma_{t}= -\frac{i}{2} \Omega \, \sigma^{\hat{x}\hat{y}},
\label{eq:Gamma}
\end{align}
where
\beqn
\sigma^{\hat{x}\hat{y}} \equiv \Sigma_z =
\left(\begin{array}{cc}
\sigma^3 & 0 \\
0 & \sigma ^3 
\end{array}\right)
\label{eq:Sigma:z}
\eeqn
in the Dirac representation of the gamma matrices:
\beqn
\gamma ^{\hat{t}}=
\begin{pmatrix}
\bbbone & 0 \\
0 & -\bbbone 
\end{pmatrix}
, \quad
\gamma ^{\hat{i}}=
\begin{pmatrix}
0 &\sigma_i \\
-\sigma_i &0 
\end{pmatrix}
,\quad
\gamma ^5=
\begin{pmatrix}
0 & \bbbone \\
\bbbone & 0 
\end{pmatrix}
. \qquad
\label{eq:gamma:Dirac}
\eeqn

Equation~\eq{eq:Dirac:rotating} is supplemented with the MIT boundary condition at the boundary of the cylinder:
\beqn
i \gamma^\mu n_\mu(\varphi) \psi(t,z,R,\varphi) = \psi(t,z,R,\varphi)\,.
\label{eq:MIT:bc}
\eeqn
where the spatial vector $n_\mu(\varphi) = (0, R \cos\varphi, - R \sin\varphi, 0)$ is normal to the cylinder surface. This condition ``confines'' the fermions inside the cavity by enforcing the normal component $j_{\bs n} \equiv - j^\mu n_\mu$ of the fermionic current $j^\mu = {\bar \psi} \gamma^\mu \psi$ to vanish at the surface of the cylinder $j_{\bs n}(\rho = R) = 0$.

\subsection{Bulk states}

A general solution of the Dirac equation~\eq{eq:Dirac:rotating} in the (co)rotating reference frame~\eq{eq:varphi} has the following form:
\beqn
U^\lambda_j = \frac{1}{2\pi} e^{- i \tE t + i k z} u^\lambda_j(\rho,\varphi)\,,
\label{eq:U:j}
\eeqn
where $u^\lambda_j$ is an eigenspinor characterized by the eigenstate helicity $\lambda = \pm 1/2$, the $z$-component of momentum $k \equiv k_z \in \R$, the projection of the quantized angular momentum $m \equiv m_z \in \Z$ onto the $z$ axis, and the radial quantum number $l = 1,2,\dots$ which describes the behavior of the solution in terms of the radial $\rho$ coordinate. The helicity $\lambda$ of the state is the eigenvalue of the helicity operator ${\hat W} = {\bs \hat P} \cdot \bs{\hat J} / p$,
\beqn
{\hat W} U^\lambda_{E k_z m} = \lambda U^\lambda_{E k_z m}\,,
\eeqn
where ${\bs \hat P} = - i {\bs \partial}$ is the momentum operator and ${\bs \hat J}$ is the angular momentum operator. In the absence of magnetic field the helicity operator $\hat W$ has the following simple form:
\beqn
 {\hat W }= \begin{pmatrix}
  {\hat h} & 0 \\
  0 & {\hat h}
 \end{pmatrix}, \qquad
 {\hat h} = \frac{{\bs \sigma} \cdot {\bs {\hat P}}}{2p},
\label{eq:hat:h}
\eeqn
where $p \equiv \sqrt{E - M^2} > 0$ is the magnitude of the spatial momentum defined as follows: 
\beqn
{\bs {\hat P}}^2 U^\lambda_j = p^2 U^\lambda_j\,.
\eeqn
Here the notation $j = (k_z, m, l)$ is used to denote a set of quantum numbers~\cite{Ambrus:2015lfr}. 

The energy in the corotating frame $\tE_j$ is related to the energy $E_j$ in the laboratory frame as follows:
\beqn
\tE_j = E_j - \Omega \Bigl(m + \frac{1}{2}\Bigr) \equiv E_j - \Omega \mu_m\,,
\label{eq:Energy}
\eeqn
where $\mu_m$ can be identified with the quantized value of the $z$-component of the total angular momentum 
\beqn
{\hat J}_z \psi = \mu_m \psi\,, \qquad \mu_m =  m + \frac{1}{2}\,,
\label{eq:mu:m}
\eeqn
which comprises the orbital and spin parts:
\beqn
\hat{J}_z  =
- i\partial_\varphi + \frac{1}{2}
\Sigma_z
,
\label{eq:hat:J}
\eeqn
where the matrix $\Sigma_z$ is given in Eq.~\eq{eq:Sigma:z}.

The solutions of the Dirac equation which satisfy the MIT boundary conditions~\eq{eq:MIT:bc} are linear combinations of positive and negative helicity spinors:
\begin{equation}
U^{\text {MIT}}_{j} = {\mathcal {C}}^{\text {MIT}}_{j} \left[ {\bb} \, U_{Ekm}^{+} + U_{Ekm}^{-} \right],
\label{eq:U:spinor:MIT}
\end{equation}
where the four-spinors with a definite helicity $\lambda$
\begin{equation}
 u^\lambda_j(\rho, \varphi) = \frac{1}{\sqrt{2}}
\left(
\begin{array}{l}
  \bE_+ \phi^\lambda_j\\
  \frac{2\lambda E}{|E|} \bE_- \phi^\lambda_j
\end{array}
\right)
\label{eq:u}
\end{equation}
are expressed with the two-spinors
\beqn
 \phi^\lambda_j(\rho, \varphi) = \frac{1}{\sqrt{2}}
 \begin{pmatrix}
 \bp_{\lambda }e^{im\varphi} J_m(q\rho /R)\\
  2i\lambda \bp_{-\lambda} e^{i(m+1)\varphi} J_{m+1}(q\rho/R)
 \end{pmatrix},
\label{eq:two:spinors}
\eeqn
which are eigenspinors of the two-component helicity operator $\hat h$~\eq{eq:hat:h}:
\beqn
	\begin{pmatrix}
	k_j & \hat{P}_- \\
	\hat{P}_+  & -k_j
	\end{pmatrix}
\frac{\phi_j (\rho, \phi) }{2p_j}
= \lambda_j \phi_j (\rho, \phi)\qquad
\label{eq:h:phi}
\eeqn
with $\hat{P}_{\pm} = \hat{P}_x \pm i \hat{P}_y = -ie^{\pm i\varphi}\left(\partial_{\rho} \pm i \rho ^{-1}\partial_{\varphi} \right)$.

In the eigenfunctions~\eq{eq:U:spinor:MIT} the degree of mixing between positive and negative helicity states is determined by the parameter
\beqn
\bb = \frac{\bE_+ \bp_+ + \bE_- \bp_-\, \jj_{ml} \, \sign{E}}{\bE_+ \bp_- + \bE_- \bp_+\, \jj_{ml}  \, \sign{E}},
\eeqn
where 
\beqn
 \bp_\pm \equiv \bp_{\pm 1/2} & = & \sqrt{1 \pm \frac{k_z}{p}}, 
 \nonumber \\
 \bE_\pm \equiv \bE_{\pm 1/2} & = & \sqrt{1 \pm \frac{M}{E}},
\eeqn
are, respectively, the momentum- and energy-related quantities which depend explicitly on the helicity of the eigenmodes, and 
\beqn
p= \sqrt{k^2_z + \frac{q^2}{R^2}}
\eeqn
is (the modulus of) the effective momentum which incorporates the longitudinal continuous momentum $k_z$ and the transverse (radial) discrete momentum number $q \equiv q_{ml}$. We also used the notation~\cite{Ambrus:2015lfr}
\beqn
\jj_{ml} = \frac{J_m(q_{ml})}{J_{m+1}(q_{ml})},
\eeqn
where $J_m(x)$ is the Bessel function.

The dimensionless real-valued and positive quantity $q_{ml}$ is the $l^{\mathrm{th}}$ real-valued positive root $(l=1,2, \dots )$ of the following equation:
\beqn
J_{m}^2(q) + \frac{2 M R}{q} \,J_m(q) J_{m+1}(q) -  J^2_{m+1}(q)  = 0\,.
\label{eq:J}
\eeqn

The normalization coefficient
\begin{widetext}
\beqn
& &  {\mathcal {C}}_{j}^{\text{MIT}} = \frac{1}{R \left|J_{m+1}(q_{m,l}R)\right|} 
\cdot \sqrt{\frac{\bp_-^2 + \bp_+^2 \, \jj_{ml}^2}
 {(\jj_{ml}^2 + 1)(\jj_{ml}^2 - (2m + 1) \frac{\jj_{ml}}{q_{m,l}R} + 1) - (\jj_{ml}^2 - 1)\frac{\jj_{ml}}{q_{m,l}R}}} \,.
\eeqn
\end{widetext}
ensures that these modes are orthonormalized
\beqn
\avr{U^{\text{MIT}}_{j}, U^{\text{MIT}}_{j'}} = \delta(k_j - k_{j'}) \delta_{m_j, m_{j'}} \delta_{l_j, l_{j'}} \theta(E_j E_{j'}),\qquad 
\label{eq:normalization}
\eeqn
with respect to the inner Dirac product:
\beqn
\avr{\psi, \chi} = \int_{-\infty}^{+\infty} d z \int_0^{2\pi} d \varphi \int_0^R d \rho \, \rho \, \psi^\dagger(x) \chi(x)\,.
\eeqn

The energies of the eigenmodes in the laboratory frame are as follows:
\beqn
E_j \equiv E_{ml}(k_z,M) = \pm \sqrt{k_z^2 + M^2 + \frac{q_{ml}^2}{R^2}}\,,
\label{eq:E:j}
\eeqn
where the plus (minus) sign corresponds to the particle (antiparticle) modes.

The density $\bar\psi\gamma^0\psi \equiv \psi^\dagger \psi$ of the wavefunctions~\eq{eq:u} is not localized at the boundary of the cylinder, and therefore we refer to these solutions as to the ``bulk eigenmodes''. They should be discriminated from the ``edge'' solutions (to be discussed below) for which the density is concentrated at the boundary of the cylinder. From Eq.~\eq{eq:E:j} we conclude that the masses of the bulk states $M^\bulk$ defined as 
\beqn
M^\bulk_{ml} = \sqrt{M^2 + \frac{q_{ml}^2}{R^2}}\,
\label{eq:M:bulk}
\eeqn
are higher than or equal to the mass of the fermion $M$.

The reflection $m \to - 1 - m$, corresponding to the sign flips of the total angular momentum~\eq{eq:mu:m} $\mu_m \to - \mu_m$, leaves the $q_{ml}$  solutions unchanged, 
\beqn
q_{ml} \to q_{-1-m,l} \equiv q_{ml}\,.
\label{eq:qml:flips}
\eeqn
This property implies that the mass spectrum~\eq{eq:M:bulk} and, consequently, the energy spectrum of the bulk modes is invariant under the flips $\mu_m \to - \mu_m$.

\begin{figure}[!thb]
\includegraphics[scale=0.55,clip=true]{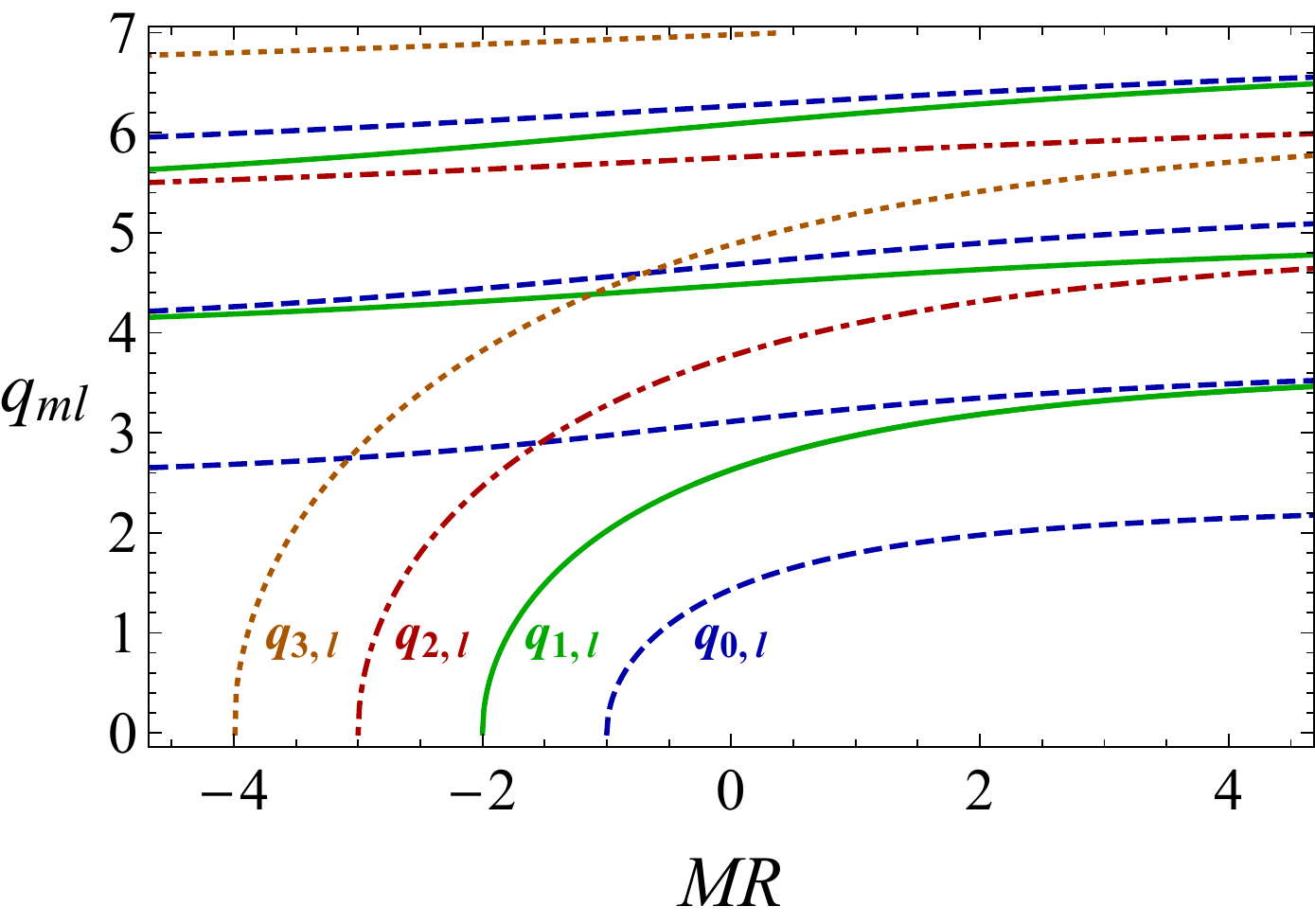}\\
\hskip 8mm (a)\\[3mm]
\includegraphics[scale=0.55,clip=true]{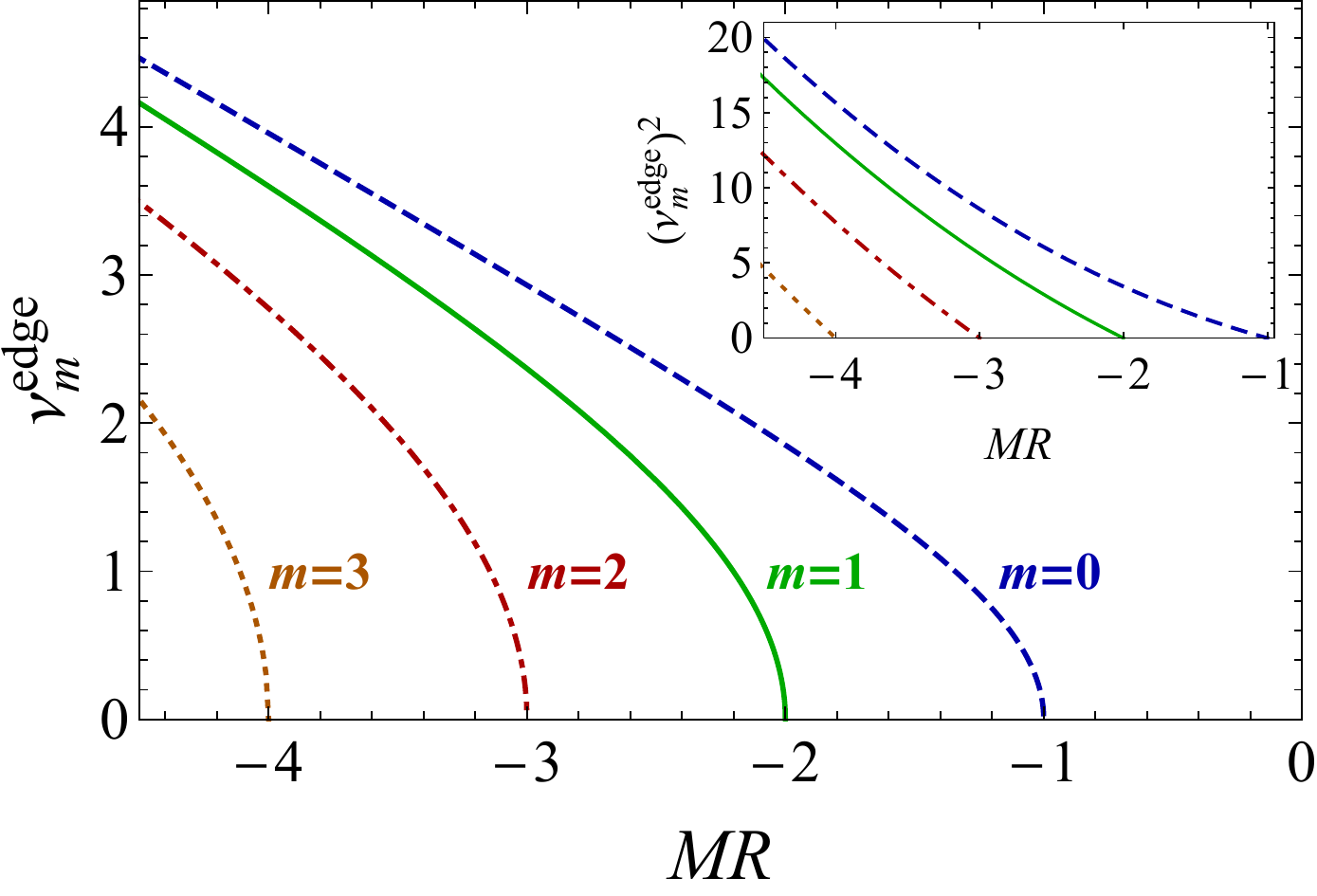} \\
\hskip 8mm (b)  
\caption{(a) Solutions of the eigenvalue equation~\eq{eq:J} as the function of the fermion mass $M$: (a) the real-valued solutions $q \geqslant 0$ corresponding to the bulk modes (from Ref.~\cite{Chernodub:2016kxh}) and (b) the purely imaginary solutions $q^\ed = i \nu^\ed$ with $\nu \geqslant 0$ corresponding to the edge modes. The inset shows $(\nu^\ed)^2$~vs~$M$.}
\label{fig:qmls}
\end{figure}

A couple of real-valued solutions $q_{ml}$ of Eq.~\eq{eq:J} are shown in Fig.~\ref{fig:qmls}(a) as a function of the fermion mass $M$. As the mass $M$ decreases, the lowest ($l=1$) real-valued modes $q_{ml} \geqslant 0$ touch the $q=0$ axis and disappear one by one at the critical values of the (negative) fermion mass:
\beqn
M_c^{(m)} = 
- \frac{1}{R} \left( |\mu_m| + \frac{1}{2} \right) \equiv
\left\{ 
\begin{array}{cll}
- \frac{1+m}{R}\,, & \quad &  m \geqslant 0,\\[2mm]
 \frac{m}{R}\,, & \quad &  m < 0.
\end{array}
\right. \qquad
\label{eq:M:c}
\eeqn
As the values $q_{m,1}$ and $q_{-1-m,1}$ coincide with each other due to the reflection invariance~\eq{eq:qml:flips}, the real-valued $q$ will disappear in pairs at the critical mass points~\eq{eq:M:c}. Contrary to the ground state with $l=1$, the excited $l \geqslant 1$ bulk states do not disappear from the spectrum. 

Finally, we would like to stress that the values of the critical mass~\eq{eq:M:c} depend on the type of the boundary condition at the boundary of the cylinder. For example, if we flip the sign of the vector $n_\mu$ in the MIT boundary condition~\eq{eq:MIT:bc}, then the mass critical values~\eq{eq:M:c} would also change the sign, $M_c^{(m)}(-n_\mu) = - M_c^{(m)}(n_\mu)$ so that disappearance of the ground state ($l=1$) modes would then happen at the positive fermion masses, $M_c^{(m)}{>}0$. With the more general chiral boundary conditions parameterized by the chiral angle $\Theta$~\cite{Lutken:1983hm},
\beqn
\bigl[i \gamma^\mu n_\mu(\varphi) - e^{- i \Theta \gamma^5} \bigr] \psi(t,z,\rho,\varphi) {\biggl |}_{\rho = R} = 0\,,
\label{eq:chiral:theta}
\eeqn
the critical masses becomes as follows~\cite{Chernodub:2017ref}:
\beqn
M_c^{(m)}(
\Theta) = 
\frac{M_c^{(m)}(0)}{\cos \Theta}
\equiv
\left\{ 
\begin{array}{cll}
- \frac{1+m}{\cos \Theta} \frac{1}{R}\,, & \quad &  m \geqslant 0,\\[2mm]
 \frac{m}{\cos \Theta} \frac{1}{R}\,, & \quad &  m < 0.
\end{array}
\right.\qquad
\label{eq:M:c:theta}
\eeqn
In particular, at the specific values of the chiral angle $\Theta=\pm \pi/2$ the ground state levels never disappears.

\subsection{Edge states}

Besides the bulk eigenfunctions with real-valued solution $q=q_{ml}$ the system contains also quite peculiar eigenstates which are localized at the boundary of the cylinder. These are the edge states which correspond to  purely imaginary solutions of  Eq.~\eq{eq:J}:
\beqn
q_m^{\mathrm{edge}} = i \nu_m^{\mathrm{edge}},
\label{eq:q:nu:edge}
\eeqn 
with a real $\nu_m^{\mathrm{edge}} \geqslant 0$.\footnote{As in the case of the bulk modes, the solutions $\nu_m^{\mathrm{edge}}$ and $-\nu_m^{\mathrm{edge}}$ correspond to the same eigenmode.} 

Using the relation $J_m(i x) = i^m I_m(x)$ we get from Eq.~\eq{eq:J} the following equation which determines $\nu$:
\beqn
I_{m}^2(\nu) + \frac{2 M R}{\nu} \,I_m(\nu) I_{m+1}(\nu) + I^2_{m+1}(\nu)  = 0, \qquad
\label{eq:I}
\eeqn
where $I_m(x)$ is the modified Bessel function.

In Fig.~\ref{fig:qmls}(b) we show the solutions of Eq.~\eq{eq:I} as the function of the fermion mass $M$. First of all we notice that there is only one edge eigenmode for each value of the orbital momentum~$m$. Moreover, the edge modes $\nu^\ed_m$ appear at the critical mass points~\eq{eq:M:c} where the lowest bulk modes $q_{m,1}$ disappear (as the fermion mass $M$ diminishes). Therefore we conclude that at the critical mass points~\eq{eq:M:c} the lowest bulk modes~\eq{eq:M:c} are transformed into the edge modes and vise versa.

The energy $E$ of the edge states in the laboratory frame is as follows:
\beqn
E^\ed_m  = \pm \sqrt{p^2 + M^2} \equiv \pm \sqrt{k^2 + (M^\ed_m)^2}\,,
\label{eq:E:edge}
\eeqn
where
\beqn
p = \sqrt{k^2 - \frac{\nu^2_m}{R^2}}\,,
\label{eq:p:edge}
\eeqn
is an analogue of momenta. The plus (minus) sign in Eq.~\eq{eq:E:edge} corresponds to the particle (antiparticle) modes similarly to the bulk modes~\eq{eq:E:j}.

Equation~\eq{eq:E:edge} implies that contrary to the masses of the bulk states~\eq{eq:M:bulk} the masses of the edge states are smaller or equal to the mass of the fermion $M$:
\beqn
M^\ed_m = \sqrt{M^2 - \frac{\nu^2_m}{R^2}}\,.
\label{eq:M:edge}
\eeqn
Notice that due to the inequality $|\nu_m| < M R$ the masses~\eq{eq:M:edge} of the edge states and their energies $E^\ed$ always remain real numbers while the effective momentum $p$ may take become purely imaginary for longitudinal momenta $|k| < \nu/R$. In other words, for the edge modes $\nu^2>0$, $k^2 >0$ and $E^2 >0$ while $p^2$ may take both positive and negative values.

In the rotating frame the energy of the edge mode follows from Eq.~\eq{eq:Energy}:
\beqn
{\widetilde E}^\ed_m  = E^\ed_m  - \Omega \mu_m\,.
\label{eq:E:edge:rotating}
\eeqn
where $E^\ed_m$ is the energy of the edge modes in the laboratory frame~\eq{eq:E:edge}.

\begin{figure}[!thb]
\includegraphics[scale=0.55,clip=true]{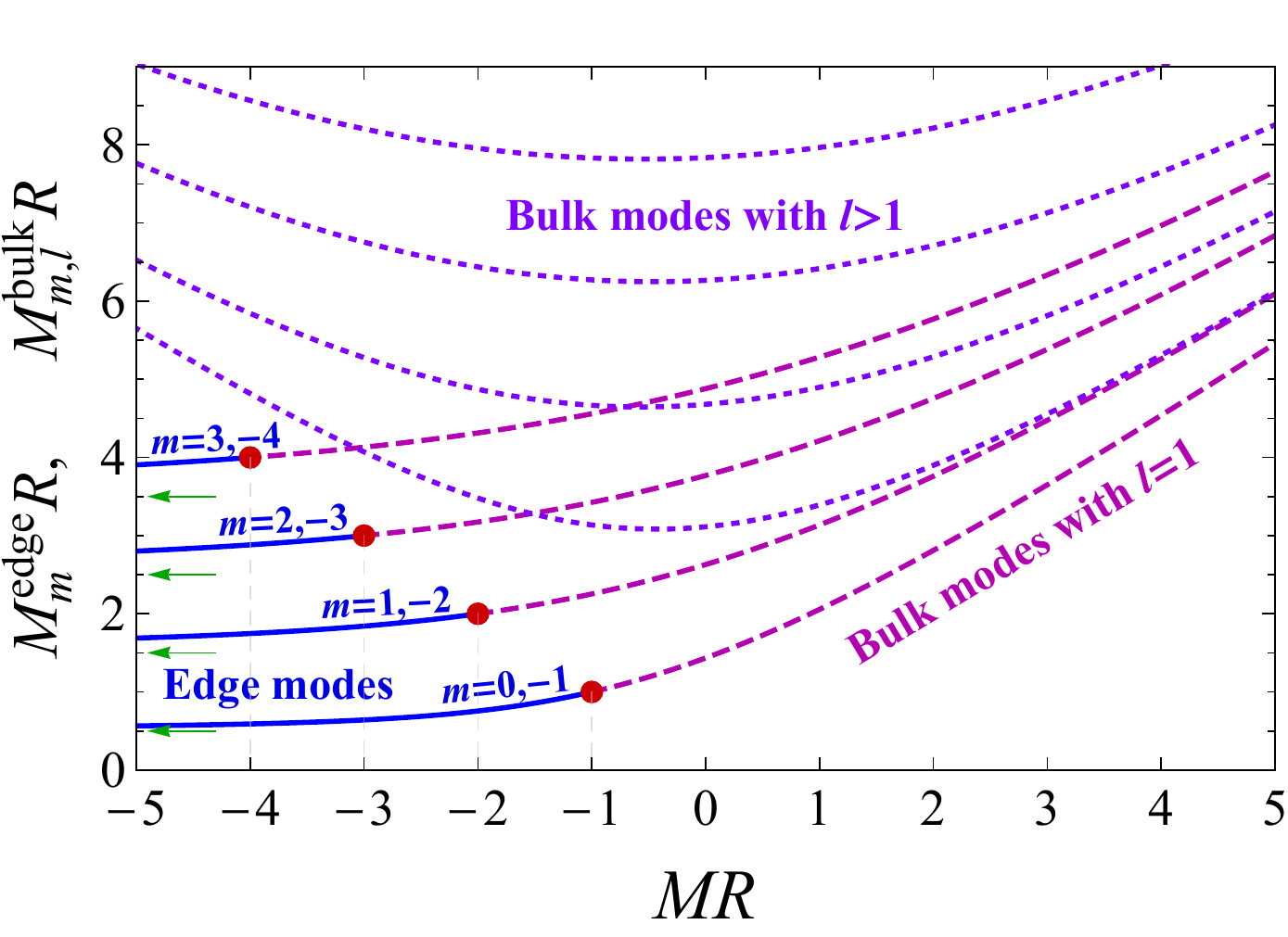} 
\caption{The masses~\eq{eq:M:edge} of the edge states (the solid blue lines) and the masses~\eq{eq:M:bulk} of the lowest ($l=1$) bulk states (the dashed magenta lines) as the function of the fermion mass $M$ in the absence of magnetic field, $B=0$. Four lowest states are shown. The critical points~\eq{eq:M:c} are marked by the red dots (and the thin gray lines). The asymptotic masses of the edge states~\eq{eq:M:edge:infty} in the limit $M \to - \infty$ are shown by the green arrows. Four lowest $l>1$ bulk states are shown by the dotted lines.}
\label{fig:masses}
\end{figure}

In Fig.~\ref{fig:masses} we show the mass spectrum both for the bulk modes~\eq{eq:M:bulk} and for the edge modes~\eq{eq:M:edge}. This figure clearly demonstrates that the ground state $l=1$ becomes the edge mode as the critical point~\eq{eq:M:c} is passed for each fixed $m$. 

Similarly to the bulk modes~\eq{eq:qml:flips}, a reflection in the sign of the total angular momentum~\eq{eq:mu:m}, $\mu_m \to - \mu_m$, leaves the $\nu_{ml}$ eigenvalues unchanged, 
\beqn
q_{ml} \to q_{-1-m,l} \equiv q_{ml}\,.
\label{eq:numl:flips}
\eeqn
Therefore the energy spectrum of the edge modes is symmetric with respect to the flips $\mu_m \to - \mu_m$. Both bulk and edge modes are degenerate in the absence of external magnetic field.

The two-spinors of the edge eigenmode with definite helicity $\lambda$ is given the spinor
\beqn
\phi_\lambda = C_\phi {\check \phi}_\lambda\,, 
\quad 
{\check \phi}_\lambda = \left(  
\begin{array}{c}
(k+2 p \lambda) e^{i m \varphi} I_m(\nu \rho/R)\\
- i \nu e^{i (m+1) \varphi} I_{m+1}(\nu \rho/R)\\
\end{array}
\right),\quad
\label{eq:phi:edge}
\eeqn
where $C_\phi$ is a normalization constant and we implied that the check mark over a spinor means that this spinor is not normalized.

In order to prove that the spinor~\eq{eq:phi:edge} is the eigenmode of the helicity operator~\eq{eq:h:phi} we used the following useful relations:
\beqn
& & \hat{P}_+ \left[e^{im\varphi} I_m\left(\nu\frac{\rho}{R}\right)\right] = \frac{\nu}{i R} e^{i (m+1)\varphi} I_{m+1}\left(\nu\frac{\rho}{R}\right)\,,\\
& & \hat{P}_-\left[e^{i(m+1)\varphi} I_{m+1}\left(\nu\frac{\rho}{R}\right)\right] = \frac{\nu}{i R}  e^{i m \varphi} I_m\left(\nu\frac{\rho}{R}\right)\,,
\eeqn
for the operators
\beqn
\hat{P}_{\pm} = -ie^{\pm i\varphi}\left(\partial_{\rho} \pm i \rho ^{-1}\partial_{\varphi}\right)\,.
\eeqn

The two-spinors for the bulk modes~\eq{eq:two:spinors} were normalized using the condition~\cite{Ambrus:2015lfr}:
\beqn
\sum_{m \in Z} \phi^{\lambda,\dagger}_{Ekm}\phi^{\lambda}_{Ekm} = 1\,,
\eeqn
which utilized the convenient summation property of the Bessel functions:
\beqn
\sum_{m \in \Z} J_m^2(x) = 1.
\eeqn
The edge states~\eq{eq:phi:edge} depend on modified, rather then usual, Bessel functions that possess a different summation rule:
\beqn
\sum_{m \in \Z} (-1)^m I_m^2(x) = 1.
\label{eq:sum:I}
\eeqn
This equations suggests that the edge eigenmodes~\eq{eq:phi:edge} should be normalized according to another normalization relation:
\beqn
\sum_{m \in Z} (-1)^m \phi^{\ed,\dagger}_{Ekm} \phi^{\ed}_{Ekm} = 1\,,
\label{eq:normalization:edge}
\eeqn
which has a less clear physical sense. Nevertheless, for the sake of completeness, we give the value of the prefactor $C_\phi$ corresponding to the normalization~\eq{eq:normalization:edge}:
\beqn
C_\phi & = & \frac{1}{\sqrt{2}} \frac{1}{\sqrt{k^2 + 2 \lambda k {\mathrm {Re}}\, p + {\mathrm {Im}}^2 p}} \nonumber \\
& = &
\frac{1}{\sqrt{2}}  \cdot \left\{
\begin{array}{lll}
{(k^2 + 2 \lambda k p)}^{-1/2}, & \quad & p^2 > 0,\\[1mm]
1/\nu, & \quad & p^2 < 0.
\end{array}
\right.
\eeqn

In the corotating reference frame the Dirac equation, if expressed via the corotating coordinates, has the same form as the standard Dirac equation in the laboratory frame in the absence of rotation. Using the explicit representation of the $\gamma$ matrices~\eq{eq:gamma:Dirac} the Dirac equation~\eq{eq:Dirac:rotating} in the corotating frame can be rewritten as follows
\beqn
\left( i \dirac - M \right) U^\ed_{j,\lambda} & \equiv &
\left(
\begin{array}{cc}
E - M  & - 2 p {\hat h} \\
2 p {\hat h} & - (E+M)
\end{array}
\right) 
U^\ed_{j,\lambda} =0, \notag
\eeqn
or
\beqn
\left(
\begin{array}{cc}
E - M  & - 2 p \lambda \\
2 p \lambda & - (E+M)
\end{array}
\right) 
\Psi^\ed_{j,\lambda}  = 0,
\label{eq:Dirac:edge:1}
\eeqn
where we set
\beqn
U^\ed_{j,\lambda}&=&\frac{1}{2\pi}e^{-i{\widetilde E}_j^{\mathrm{edge}}t+ik_jz}\Psi^\ed_{j,\lambda}, \notag \\
\Psi^\ed_{j,\lambda}&= &
\left(
\begin{array}{c}
C_{\mathrm{up}} \,  {\check \phi}^\lambda_j \\[1mm]
C_{\mathrm{down}} \, {\check \phi}^\lambda_j
\end{array}
\right)\,,
 \label{eq:Psi:j:lambda}
\eeqn
and then used the fact that the two-spinors ${\check \phi}^\lambda_j$, Eq.~\eq{eq:phi:edge}, are the eigenfunctions of the helicity operator $\hat h$, Eq.~\eq{eq:hat:h}.

The self-consistency of the Dirac equation for the edge modes~\eq{eq:Dirac:edge:1} gives us the expression for their energy~\eq{eq:E:edge} and fixes the coefficients $C_{\mathrm{up}}$ and $C_{\mathrm{down}}$ in Eq.~\eq{eq:Psi:j:lambda} up to the overall normalization factor (set to unity in this expression):
\beqn
{\check \Psi}^\ed_{j,\lambda} = \left(
\begin{array}{c}
(E+M){\check \phi}^\lambda_j \\
2 \lambda p \, {\check \phi}^\lambda_j
\end{array}
\right)\,,
\label{eq:Psi:j:sol}
\eeqn

Denoting $\Psi^\ed_j = (\Psi_\uparrow, \Psi_\downarrow)^{T}$, the MIT boundary conditions~\eq{eq:MIT:bc} may be explicitly written as follows:
\beqn
(i \slashed{n} -1) \Psi^\ed_j = 
- \left(
\begin{array}{cc}
\bbbone & i \sigma^\rho \\
- i \sigma^\rho  &  \bbbone
\end{array}
\right) 
\left(
\begin{array}{c}
\Psi_\uparrow \\
\Psi_\downarrow
\end{array}
\right) = 0 \,,
\label{eq:MIT:components}
\eeqn
where we set $\rho = R$ and defined
\beqn
\sigma^{\rho}= \sigma_1\cos \varphi + \sigma_2\sin\varphi\,.
\label{eq:sigma:rho}
\eeqn

The four-spinor solutions satisfying these conditions should involve both $\lambda = \pm 1/2$ helicities~\cite{Ambrus:2015lfr}:
\beqn
\Psi^\ed_j & \equiv & \Psi^\ed_{j,{\mathrm{MIT}}} = \sum_{\lambda = \pm}
C_j^\lambda {\check \Psi}^\ed_{j,\lambda}
\label{eq:Psi:j}
\nonumber \\
& \equiv & 
\left(
\begin{array}{c}
(E+M) \left(C_j^+ {\check \phi}^+_j + C_j^- {\check \phi}^-_j \right) \\
p \, \left(C_j^+ {\check \phi}^+_j - C_j^- {\check \phi}^-_j \right)
\end{array}
\right).
\eeqn
because the MIT boundary condition~\eq{eq:MIT:bc} breaks the helicity conservation.

The self-consistency requirement for the MIT condition~\eq{eq:MIT:components} and \eq{eq:Psi:j} gives us the relation~\eq{eq:I} which determines the value of the parameter~$\nu$.

From Eq.~\eq{eq:I} it follows that the nontrivial solutions for $\nu = \nu_m$ exist if and only if $M< 0$. Solving Eq.~\eq{eq:I} as a quadratic equation we get:
\beqn
i_m \equiv \frac{I_{m+1}(\nu_m)}{I_m(\nu_m)} & = & - \frac{M R + \sign{\mu_m} M^\ed_m R}{\nu_m} 
\label{eq:II:ratio} \nonumber \\
& \equiv &
\left\{
\begin{array}{lll}
- \frac{M R + M^\ed_m R}{\nu_m}, & \quad & m \geqslant 0,\\[1mm]
- \frac{M R - M^\ed_m R }{\nu_m}, & \quad & m < 0,
\end{array}
\right.
\eeqn
where the angular momentum $\mu_m$ and the mass of the edge state $M^\ed_m$ are given in Eqs.~\eq{eq:mu:m} and \eq{eq:M:edge}, respectively.

The coefficients in Eq.~\eq{eq:Psi:j} satisfy the relation:
\beqn
\sum_{\lambda=\pm 1/2} C^\lambda_j (1 + 2 \lambda \kappa) = 0\,,
\label{eq:C:lambda}
\eeqn
where 
\beqn
\kappa_m(k) 
& = & \frac{p\left[E_m(k)+M+ \nu i_m/R\right]}{k[E_m(k)+M]} \nonumber \\
& = & \frac{p \left[E_m(k) - \sign{\mu_m} M^\ed_m \right]}{k\left[E_m(k)+M \right]} \nonumber \\
& \equiv & \frac{k\left[E_m(k) - M\right]}{p \left[E_m(k) + \sign{\mu_m}  M^\ed_m\right]}\,,
\eeqn
and we adopted the usual convention $C_j^{\pm 1/2} \equiv C_j^{\pm}$. One can also rewrite the last expression in the following explicit form:
\beqn
\kappa_m(k) = \frac{p \left(\sqrt{k^2 + M^2 - \frac{\nu_m^2}{R^2}} - \sign{\mu_m} \sqrt{M^2 - \frac{\nu_m^2}{R^2}}\right)}{k \left(\sqrt{k^2 + M^2 - \frac{\nu_m^2}{R^2}}+M \right)} \,.
\nonumber
\eeqn

Combining \eq{eq:C:lambda} and \eq{eq:Psi:j} we get the edge eigenmode in the explicit form:
\beqn
\Psi_j = C_0 
\left(
\begin{array}{l}
(E+M) (\kappa k - p) e^{i m \varphi} I_m\left( \nu_m \frac{\rho}{R} \right) \\[2mm]
\frac{\nu_m}{i R} (E+M) \kappa  e^{i (m+1) \varphi} I_{m+1}\left( \nu_m \frac{\rho}{R} \right)\\[2mm]
p(p\kappa - k) e^{i m \varphi} I_m\left( \nu_m \frac{\rho}{R} \right) \\[2mm]
i p \frac{\nu_m}{R}  e^{i (m+1) \varphi} I_{m+1}\left( \nu_m \frac{\rho}{R} \right)
\end{array}
\right).
\label{eq:psi:edge:modes}
\eeqn

\begin{widetext}
The overall constant $C_0$ is determined by the orthonormalization condition given in Eq.\eq{eq:normalization}. For the edge mode, the Dirac inner product is given by
\beqn
\left<U_j ^\ed , U_{j'}^\ed \right> &=& \delta\left(k-k'\right)\delta_{m m'}\theta(E_j E_{j'})|C_0|^2 \notag \\
&\times& \Biggl[ \left\{(E_j+M)^2(\kappa_m k -p)^2 + p^2(p\kappa_m-k)^2 +\left(\frac{\nu_m^2}{R^2}(E_j+M)^2\kappa_m ^2 + p^2 \frac{\nu_m^2}{R^2}\right) \right\}\mathcal{I}_{m+1/2}^+ \notag \\
&+&\left\{(E_j+M)^2(\kappa_m k -p)^2 + p^2(p\kappa_m-k)^2 -\left(\frac{\nu_m^2}{R^2}(E_j+M)^2\kappa_m ^2 + p^2 \frac{\nu_m^2}{R^2}\right)\right\}\mathcal{I}_{m+1/2}^- \Biggr],
\eeqn
where $\mathcal{I}_{m+1/2}^\pm$ is defined as 
\beqn
\mathcal{I}_{m+1/2}^+(\nu_m) = \int_0^R d \rho \, \rho\, \frac{I_m^2(\nu_m\frac{\rho}{R}) +I_{m+1}^2(\nu_m\frac{\rho}{R})}{2} & = & \frac{R^2}{2}\frac{1}{\nu_m} I_m(\nu_m) I_{m+1}(\nu_m) , \notag \\
\mathcal{I}_{m+1/2}^-(\nu_m) = \int_0^R d \rho \, \rho\, \frac{I_m^2(\nu_m\frac{\rho}{R}) -I_{m+1}^2(\nu_m\frac{\rho}{R})}{2} & = & \frac{R^2}{2} \Bigl[ I_m^2(\nu_m) - \frac{2 m+1}{\nu_m} I_m(\nu_m) I_{m+1}(\nu_m) -  I_{m+1}^2(\nu_m)\Bigr].
\eeqn
Thus, the normalization coefficient $C_0$ is given by the following expression:
\beqn
C_0 &=& \frac{1}{|I_m(\nu_m)|}\frac{\sqrt{2}k}{p\sqrt{\nu_m}}\Biggl[\left[\left\{k^2+{(E-M)^2}+\frac{\nu^2}{R^2}\right\}i_m^2+4\frac{\nu M}{R}i_m+\left\{k^2+{(E+M)^2}+\frac{\nu^2}{R^2}\right\}\right]i_m \notag \\
&+&\left[2E(E-M)i_m^2-4\frac{\nu E}{R}i_m-2E(E+M)\right]\left(1-\frac{2m+1}{\nu_m}i_m-i_m^2\right)
\Biggr]^{-1/2}.
\eeqn
\end{widetext}

The special case $k \equiv k_z = 0$ one gets the following explicit expression of the edge eigenmode:
\beqn
\Psi_j {=} C_0 
\left(
\begin{array}{l}
\theta(\mu_m) \left( M + M^\ed_m \right) e^{i m \varphi} I_m\left( \nu_m \frac{\rho}{R} \right)\\[2mm]
\theta(- \mu_m) \left( M + M^\ed_m \right) e^{i (m+1) \varphi} I_m\left( \frac{\nu_m \rho}{R} \right)\\[2mm]
- i \nu_m \theta(- \mu_m) e^{i m \varphi} I_m\left( \nu_m \frac{\rho}{R} \right) \\[2mm]
- i \nu_m \theta(\mu_m) e^{i (m+1) \varphi} I_m\left( \nu_m \frac{\rho}{R} \right)
\end{array}
\right). \qquad
\nonumber
\eeqn

The normalization coefficient (for $m \geqslant 0$ so far) is given by the following expression:
\beqn
C_0 = \frac{1}{|I_m(\nu_m)|}{\left( \gamma \left[ \nu + 2 m \gamma M^\ed - 2 \gamma \left(M^\ed\right)^2 \right] \right)}^{-\frac{1}{2}},
\nonumber
\eeqn
where
\beqn
\gamma_m = -\frac{(M + M^\ed) R}{\nu}\,.
\eeqn
Notice that $0<\gamma <1$.

\begin{figure}[!thb]
\includegraphics[scale=0.55,clip=true]{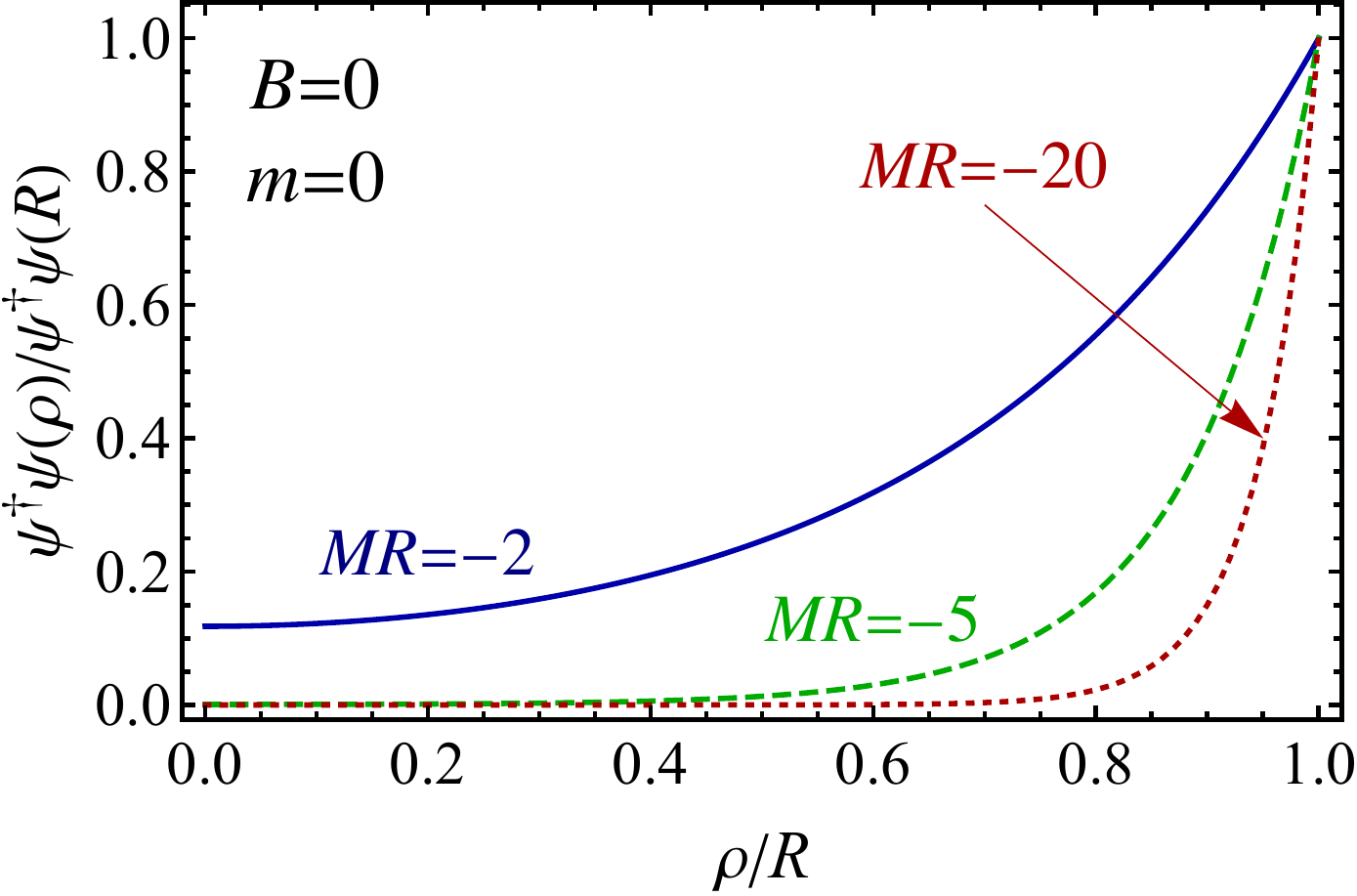} 
\caption{An example of the density of the edge modes for $k=0$ and $m=0$ at various fermionic masses $M$ in the absence of magnetic field ($B=0$).}
\label{fig:psi2}
\end{figure}

The solutions~\eq{eq:psi:edge:modes} correspond to the edge modes because their density $\bar \psi \gamma^0 \psi \equiv \psi^\dagger\psi$ grows exponentially as one approaches the edge of the cylinder at $\rho = R$, Fig.~\ref{fig:psi2}. Since all modified Bessel functions $I_n$ grow exponentially at large values of its argument, the localization length of the edge states~\eq{eq:psi:edge:modes} at the boundary of the cylinder is determined by the length scale 
\beqn
\xi_m^\ed = \frac{R}{\nu_m}\,.
\label{eq:xi:edge}
\eeqn
Thus, the edge modes are characterized by two dimensionful parameters, their mass~\eq{eq:M:edge} and the localization length~\eq{eq:xi:edge}. Notice that the former may be expressed via the latter:
\beqn
M^\ed_m = \sqrt{M^2 - {\bigl( \xi_m^\ed \bigr)}^{-2}}\,, 
\label{eq:M:edge:xi}
\eeqn

Now, let us consider the behavior of the masses of the edge modes $M^\ed$ in the limit of a large fermion mass $M$. For a large positive real $z \gg 1$ the modified Bessel functions have the following asymptotic expansion:
\beqn
I_m(z) = \frac{e^z}{\sqrt{2 \pi z}} \left( 1 + \frac{1-4m^2}{8 z} + O\bigl(z^{-2}\bigr) \right).
\label{eq:Im:expansion}
\eeqn
Substituting Eq.~\eq{eq:Im:expansion} into the relation~\eq{eq:II:ratio} we get that in the limit of a large negative mass $M$ the solutions $\nu^\ed$ behave as follows:
\beqn
\nu^\ed_m = |M| R - \frac{\mu_m^2}{2 |M| R} + O\Bigl(\bigl(M R\bigr)^{-2}\Bigr)\,, 
\label{eq:nu:edge:infty}
\eeqn
where $\mu_m$ is the total angular momentum of the mode~\eq{eq:mu:m}.

Therefore, in the limit of the infinite fermionic mass the masses of the edge modes remans finite contrary to the bulk modes $M^\ed_{ml}$ which become infinitely massive in this limit~\eq{eq:M:bulk} and therefore decouple from the system. Moreover, in the limit of large (negative) fermion mass the mass spectrum of the edge modes may be computed analytically:
\beqn
M^\ed_{\infty,m} = \lim_{M \to - \infty} M^\ed_m = \frac{|\mu_m|}{R}\,.
\label{eq:M:edge:infty}
\eeqn
We find that the masses of the edge modes~\eq{eq:M:edge:infty} are (i) finite,  (ii) quantized and (iii) independent of the fermion mass $M$. According to Eq.~\eq{eq:nu:edge:infty} the localization length~\eq{eq:xi:edge}  tends to zero in this limit. The edge states are double-degenerate as the modes with opposite angular momenta ($\mu_m$ and $\mu_{-1-m} \equiv - \mu_m$) possess the same mass. We also stress that in the absence of magnetic field there are no massless edge modes in the spectrum in a cylinder of a finite radius $R$. The modes eventually become massless in the limit of a large radius $R \to \infty$.

In conclusion of this section we would like to notice that the physical particle-antiparticle interpretation of the fermionic modes in the second-quantization formalism depends on the presence of the modes for which $E {\widetilde E} < 0$. The physical meaning of such modes is ambiguous (see Refs.~\cite{Vilenkin:1980zv,Iyer:1982ah} as well as the detailed discussion in Ref.~\cite{Ambrus:2015lfr}), and therefore the absence of such modes in the spectrum makes the theory well defined. In short, the modes $E>0$ ($E<0$) in laboratory frame are interpreted as particle (antiparticle) states in the Vilenkin quantization~\cite{Vilenkin:1980zv} while the modes with ${\widetilde E}>0$ (${\widetilde E}<0$) in the corotating frame are interpreted as particle (antiparticle) states in the quantization of Iyer~\cite{Iyer:1982ah}. Both vacua are the same provided $E_j {\widetilde E}_j > 0$ for all modes. In Ref.~\cite{Ambrus:2015lfr} it was indeed found for uniformly rotating states bounded within the light cylinder (so that with $|\Omega| R < 1$) with physically reasonable boundary conditions the condition $E_j {\widetilde E}_j > 0$ is satisfied for all bulk modes~\cite{Ambrus:2015lfr}, so that the rotating (Iyer) and laboratory (Vilenkin) vacua are equivalent. Below we show that the same identity is also true for the edge modes,
\beqn
E_m^\ed \widetilde{E}_m^\ed > 0\,,
\eeqn
provided they rotate within the light cylinder, $|\Omega| R < 1$.

Since the energy for $k\neq 0$ is grater than the one for $k=0$, we focus on the energy for $k=0$, 
\beqn
|E_m^\ed| = M_m^\ed.
\eeqn
 The derivative of $E_m^\ed$ with respect to $M$ is given by
\beqn
\frac{d|E^\ed_m|}{dM} = \frac{M-\frac{\nu_m}{R^2}\frac{d\nu_m}{dM}}{|E^\ed_m|}.
\eeqn
The derivative can be also expressed via Eq.~\eq{eq:II:ratio}:
\beqn
\frac{1}{R}\frac{d\nu_m}{dM}\frac{I_{m+1}(\nu_m)}{I_m (\nu_m)}\left[ 1 + \nu_m \frac{I'_{m+1}}{I_{m+1}}-\nu_m \frac{I'_{m}}{I_{m}}\right]
& & \nonumber \\
=  -1 - \sign{\mu_m}\frac{d|E^\ed_m|}{dM}\,, & & 
\eeqn
with $I'_m (\nu_m)=  dI_m(\nu_m)/d\nu_m$.
Using the following properties of the modified Bessel functions,
\beqn
I'_m (z) &=& \frac{m}{z}I_m(z) + I_{m+1}(z), \\
I'_{m+1} (z) &=& I_m(z) -\frac{m+1}{z} I_{m+1}(z), 
\eeqn
the derivative can be rewritten as
\beqn
\frac{d|E^\ed_m|}{dM} =\frac{2|\mu_m |M -2 |E_m^\ed |MR-|E_m^\ed |}{2|\mu_m ||E_m^\ed|-2(E_m^\ed)^2R-M}
\eeqn
If there is a local minimum at $M=M_0 < -|\mu_m|-1/2$, the energy is given by
\beqn
|E_m^\ed|R = \frac{2M_0R}{1+2M_0R}|\mu_m|.
\eeqn
Due to the 
nonnegativity
of the left hand side of the above equation, the local minimum can exist only for $MR < - 1/2$. In this region, the inequality $|E_m^\ed|R > |\mu_m|$ is satisfied, and thus $E_m^\ed{\widetilde E}_m^\ed>0$ is satisfied for $\Omega R<1$ (here we take for simplicity $\Omega >0$). There is also a possibility that the minimum is achieved at ends, $M\rightarrow -|\mu_m|-1/2$ or $M \rightarrow -\infty.$ At such points, the energies are given by $E_m^\ed =  (|\mu_m| + 1/2)/R$ and $E_m^\ed = |\mu_m|/R$, respectively. Therefore, in the region of $MR \in (-\infty, -|\mu_m|-1/2)$, the relation $ E_m^\ed{\widetilde E}_m^\ed>0$ is always satisfied for the uniform rotation within the light cylinder $\Omega R<1$.

\section{Bulk and edge solutions in the magnetic field background}
\label{sec:rotating:magnetic}

In this section we derive, following the general line of the previous section, the eigenspectrum of the Dirac fermions in the background of magnetic field.

\subsection{Dirac equation in rotating spacetime in the uniform magnetic field}

In the presence of an external magnetic field parallel to the cylinder axis ${\bs B} = (0,0,B_z \equiv B)$ the Dirac equation~\eq{eq:Dirac:rotating} is modified:
\beqn
\left[ i \gamma^\mu \left(D_\mu + \Gamma_\mu \right)- M \right] \psi = 0\,,
\label{eq:Dirac:rotating:B}
\eeqn
where $D_\mu = \partial_\mu - i e A_\mu$ is the covariant derivative. In the laboratory frame the corresponding gauge field can be chosen in the symmetric form 
\beqn
A_{\hat{i}}= \left(0, \frac{By}{2}, - \frac{Bx}{2}, 0\right)\,.
\label{eq:A:B:laboratory}
\eeqn
In the corotating frame the background gauge field is as follows:
\beqn
A_{\mu}= \left(- \frac{B \Omega r^2}{2}, \frac{By}{2}, - \frac{Bx}{2}, 0\right).
\label{eq:A:B:rotating}
\eeqn

The Dirac equation~\eq{eq:Dirac:rotating:B} can be explicitly written as follows:
\begin{widetext}
\beqn
\left[i\gamma ^{\hat{t}}\left(\partial_t+y\Omega\partial_x  -x \Omega \partial_y - \frac{i}{2}\Omega \sigma^{\hat{x}\hat{y}}\right)+ i\gamma ^{\hat{x}}\left(\partial_x +\frac{ieBy}{2}\right)+i\gamma^{\hat{y}}\left(\partial_y - \frac{ieBx}{2} \right) + i\gamma^{\hat{z}}\partial_z -M\right]\psi = 0,
\label{eq:rotmagDirac}
\eeqn
\end{widetext}

As in the absence of magnetic field the eigenvectors of the Dirac equation~\eq{eq:rotmagDirac} are labeled by the eigenvalues of commuting operators $\{\hat{\widetilde H},\hat{P}_z, \hat{J}_z, \hat{W} \}$, where $\hat{\widetilde H}$ is the corotating Hamiltonian, $\hat{P}_z$ is the $z$-component of the momentum operator, $\hat{J}_z$ is the $z$-component of the total angular momentum~\eq{eq:hat:J}, and $\hat{W}$ is the helicity operator. In the presence of magnetic field these operators coincide with the ones given in Section~\ref{sec:rotating} with the substitution $\bs{\hat{P}} \to \bs{\hat{P}} + e\bs{\hat{A}}$ which accounts for the gauge invariance of these operators.
In the presence of magnetic field the corotating energy $\widetilde{E}_j$ is related to the laboratory energy $E_j$ according to Eq.~\eq{eq:Energy}.

Notice that Eq.~\eq{eq:rotmagDirac} is gauge invariant because of the identity which holds for usual $\partial_\mu$ and covariant $D_\mu$ derivatives in the corotating reference frame:
\beqn
\partial_t +y\Omega \partial_x - x \Omega \partial_y  \equiv D_t + y\Omega D_x  - x\Omega D_y\,.
\label{eq:relation}
\eeqn
Here we used the fact that in the rotating frame the gauge field~\eq{eq:A:B:rotating} acquires the compensating time component $A_0 = - B \Omega r^2/2$.

In fact, relation~\eq{eq:relation} has a much deeper sense than just a simple mathematical identity. In the absence of the magnetic background the relation between the energies in corotating and laboratory frames is given by Eq.~\eq{eq:Energy}. Since thermodynamical and mechanical properties of the system depend on the energies in the corotating (rather then laboratory) frame it is important to figure out if the relation~\eq{eq:Energy} still holds in the presence of magnetic field $B$ or not. Indeed, in order to maintain the gauge invariance the usual derivatives $\partial_\mu$ in the presence of magnetic field in all physical operators should transform to the covariant derivatives $D_\mu = \partial_\mu - i e A_\mu$. In particular, the angular momentum operator~\eq{eq:hat:J} should become as follows
\beqn
\hat{J}_z(A) & = & J_z - i e A_\varphi 
\equiv 
- i\partial_\varphi
+ \frac{1}{2}
\Sigma_z
 -  \frac{eB r^2}{2}, \qquad
\label{eq:hat:J:B}
\eeqn
where $J_z \equiv J_z(A=0)$. Therefore we could naturally expect that in the presence of magnetic field the crucial corotating-laboratory energy relation~\eq{eq:Energy} could also be modified. In order to clarify this issue we notice that the relation~\eq{eq:Energy} comes from the relation between Hamiltonians in the rotating (${\hat {\widetilde H}} = i \partial_t$) and laboratory (${\hat H} = i \partial_{\hat t}$) reference frames
\beqn
{\hat {\widetilde H}}  = {\hat H} - \Omega {\hat J}_z\,,
\label{eq:eigenE:B0}
\eeqn
which has been used so far at vanishing magnetic field.
However, in the presence of magnetic field the gauge-covariant Hamiltonian in corotating frame is given by 
\beqn
H = i D_t \equiv i \partial_t + e A_t\,,
\label{eq:H}
\eeqn
[while the Hamiltonian in the laboratory frame ${\hat H}\equiv i D_{\hat t}$ remains untouched as $A_{\hat t} \equiv 0$ according to Eq.~\eq{eq:A:B:laboratory}] so that the eigenvalue equation for the energy levels becomes as follows:
\beqn
i D_t \psi = \bigl[{\hat H} - \Omega {\hat J}_z(A)\bigr] \psi\,.
\label{eq:eigenE:B}
\eeqn
However, taking into account in the rotating frame $A_t = \Omega A_\varphi \equiv  - B \Omega r^2/2$ [used already in Eq.~\eq{eq:relation}], we arrive to the conclusion that the ``covariantization'' of the Hamiltonian~\eq{eq:H} and the covariantization of angular momentum operator~\eq{eq:hat:J:B} exactly cancel each other in Eq.~
\eq{eq:eigenE:B} and we arrive to 
\beqn
i \partial_t \psi = \Bigl({\hat H} - \Omega J_z \Bigr) \psi\,.
\label{eq:psi:reduced}
\eeqn
Next, we notice that the energy in the corotating frame enters the wavefunction as $\psi(t,{\bs x}) = \exp\{- i {\widetilde E}_j t\} \, \psi({\bs x})$ and therefore one gets from Eq.~\eq{eq:psi:reduced}:
\beqn
{\widetilde E}_j  \psi({\bs x}) = \Bigl({\hat H} - \Omega J_z \Bigr) \psi({\bs x})\,,
\eeqn
which agrees with Eq.~\eq{eq:eigenE:B0} which, in turn, leads to the relation in question~\eq{eq:Energy}. Thus we conclude that the relation~\eq{eq:Energy} between the energies in the corotating $\widetilde E$ and laboratory $E$ frames is still valid in the presence of the magnetic field background.

\subsection{Solutions}

A general solution of the Dirac equation~\eq{eq:rotmagDirac} has the following form,
\beqn
U_j (t,z,\rho,\varphi) = \frac{1}{2\pi} e^{- i \tE_j t + i k_z z} u_j(\rho,\varphi)\,,
\label{eq:U}
\eeqn
where $u_j$ is an eigenspinor. The diagonal forms of $\hat{J}_z$ and $\hat{W}$ allow us to express the eigenspinor $u_j$ as follows
\beqn
u_j(\rho,\varphi) =
   \begin{pmatrix}
   C_j^{\mathrm{up}}\phi_j (\rho, \phi) \\
   C_j^{\mathrm{down}}\phi_j (\rho, \phi) \\   
   \end{pmatrix}\,,
\eeqn
where the two-spinor
\beqn
\phi_j(\rho, \phi) = 
   \begin{pmatrix}
   e^{im_j \varphi}\chi_j^-(\rho) \\
   e^{i(m_j+1)\varphi}\chi_j^+ (\rho)
   \end{pmatrix}\,,
\eeqn
is defined via two scalar functions~ $\chi ^{\pm}_j$ of the radial coordinate $\rho$. The helicity eigenvalue equation, $\hat{W}U_j = \lambda_j U_j$, is reduced to the following relation,
\beqn
	\begin{pmatrix}
	k_j & \hat{P}_- + e\hat{A}_- \\
	\hat{P}_+ + e\hat{A}_+ & -k_j
	\end{pmatrix}
\frac{\phi_j (\rho, \phi) }{2\sqrt{E_j^2-M^2}}
= \lambda_j \phi_j (\rho, \phi),\qquad
\label{eq:helicityeq}
\eeqn 
with $\hat{P}_{\pm} + e\hat{A}_{\pm} = -ie^{\pm i\varphi}\left(\partial_{\rho} \pm i \rho ^{-1}\partial_{\varphi} \pm eB\rho/2\right)$. The equations for $\chi ^{\pm}_j$ are written as follows:
\begin{widetext}
\beqn
\left[\partial_\rho ^2 + \frac{\partial_\rho}{\rho} -\left(\frac{m_j + 1}{\rho}\right)^2+m_jeB -\frac{e^2B^2}{4}\rho^2 + \left(E_j^2 - M^2 - k_j^2\right)\right]\chi_j^+ = 0\,,& \\
\left[\partial_\rho ^2 + \frac{\partial_\rho}{\rho} -\left(\frac{m_j}{\rho}\right)^2+\left(m_j+1\right)eB -\frac{e^2B^2}{4}\rho^2 + \left(E_j^2 - M^2 - k_j^2\right)\right]\chi_j^- = 0.&
\eeqn
\end{widetext}
Using the substitution $\xi \equiv \frac{eB}{2}{\rho^2}$, the above equations are reduced, respectively, to a simpler set of relations
\beqn
\begin{array}{r}
\xi(\chi_j^+)''+(\chi_j^+)'+\left(-\frac{1}{4}\xi+\beta^+-\frac{(m+1)^2}{4\xi}\right)\chi_j^+=0, \\[3mm]
\xi(\chi_j^-)''+(\chi_j^-)'+\left(-\frac{1}{4}\xi+\beta^--\frac{m^2}{4\xi}\right)\chi_j^-=0,
\end{array}
\qquad
\eeqn
where
\beqn
\beta^\pm = \frac{2 \mu_m\mp 1}{4}+\frac{1}{2eB}\left(E_j^2-M^2-k_j^2\right),
\eeqn
and the angular momentum $\mu_m$ is given in Eq.~\eq{eq:mu:m}.

The normalizable (regular in the origin) solutions are given by the confluent hypergeometric function $\mM(a,b; z) \equiv {}_1 F_1 (a,b; z)$~\cite{landau2013quantum, Chen:2015hfc}
\begin{align}
\chi_j^+ &= \mathcal{N}_j^+ \rho^{|m_j+1|}e^{-\frac{eB}{4}\rho^2}\mM^+\,, \label{eq:phi+} \\
\chi_j^- &= \mathcal{N}_j^-\rho^{|m_j|}e^{-\frac{eB}{4}\rho^2}\mM^-\,, \label{eq:phi-}
\end{align}
where $\mM^\pm$ is defined as
\begin{align}
\mM^+ &\equiv \mM\left(a_j^+, |m_j+1|+1, \frac{eB}{2}\rho^2\right)\,, \\
\mM^- &\equiv \mM\left(a_j^-, |m_j|+1, \frac{eB}{2}\rho^2\right)\,,
\end{align}
and $a^{\pm}_j$ is defined as 
\begin{align}
a^+_j &= \frac{1}{2}\left(|m_j+1|-m_j+1\right)-\frac{1}{2eBR^2}\left(q^B_j\right)^2\,, 
\label{eq:aj:plus} 
\\
a^-_j &= \frac{1}{2}\left(|m_j|-m_j\right)-\frac{1}{2eBR^2}\left(q^B_j\right)^2\,
\label{eq:aj:minus}
\end{align}
with $q^B_j \equiv \sqrt{E_j^2-M^2-k_j^2}R$. The coefficient $\mathcal{N}_j^+ $ can be related to the coefficient  $\mathcal{N}_j^-$ by a substitution of Eq.(\ref{eq:phi+}) and Eq.(\ref{eq:phi-}) into the helicity equation Eq.(\ref{eq:helicityeq}):
\beqn
\begin{array}{rll}
\mathcal{N}_j^+  &=\frac{+i\left(E_j^2 - M^2 - k_j^2\right)}{2\left(k_j + 2\lambda_j \sqrt{E_{j}^2-M^2}\right)(m_j+1)}\mathcal{N}_j^-, \quad &m_j\ge 0 ,  \\[4mm]
\mathcal{N}_j^+  &=\frac{2i m_j}{k_j + 2\lambda_j \sqrt{E_{j}^2-M^2}}\mathcal{N}_j^-, \quad &m_j<0 . 
\end{array}
\hskip 7mm
\eeqn
The two spinors $\phi_j^\lambda$ with the helicity $\lambda$ are written as follows
\beqn
 \phi_j^\lambda(\rho, \varphi) = \alpha_j
 \begin{pmatrix}
 f^\lambda_{j-} \mM^-_j \\
 f^\lambda_{j+} \mM^+_j 
 \end{pmatrix}\,,
 \label{eq:phi:j}
\eeqn
where  $\alpha _j$ is an overall constant and the two-spinor $(  f^\lambda_{j-}  \,\, f^\lambda_{j+} )^T$ is defined as
\beqn
  \begin{pmatrix}
  f^\lambda_{j-}  \\
  f^\lambda_{j+}  
  \end{pmatrix}
  {=} \left\{ \begin{array} {ll}
\left(
\begin{array}{c} 
2(m_j + 1) G_{m_j}(\rho,\varphi) \\
2i\lambda_j p^B_j \left(\bp^B_{-\lambda}\right)^2G_{m_j+1}(\rho,\varphi)
\end{array}
\right), & m_j \ge 0,\\
\\
\left(
\begin{array}{c} 
p_j^B \left(\bp_\lambda^B\right) ^2 G_{m_j}(\rho,\varphi) \\
  4i\lambda_j m_j G_{m_j+1}(\rho,\varphi)
\end{array}
\right), & m_j < 0,\\
\end{array} \right.
\label{eq:f:j}
\eeqn
with
\beqn
G_m(\rho,\varphi) = e^{im\varphi} \rho^{|m|}e^{-\frac{eB}{4}\rho^2}\,,
\eeqn
and
\beqn
 \bp_\pm^B = \sqrt{1 \pm \frac{k_z}{p^B_j}}, 
 \qquad
  p^B_j= \sqrt{E_j^2-M^2}.\qquad
\eeqn

Next, we use the Dirac equation Eq.(\ref{eq:rotmagDirac})  determine the constraint between $C_j^{\mathrm{up}}$ and $C_j^{\mathrm{down}}$:
\beqn
	\begin{pmatrix}
	E_j-M & -2\lambda_j\sqrt{E_j^2-M^2} \\
	2\lambda_j\sqrt{E_j^2-M^2} & -E_j -M
	\end{pmatrix}
u_j (\rho, \phi) = 0,
\label{eq:constraints_C}
\eeqn 
or 
\beqn
C_j^{\mathrm{up}} = \frac{\sqrt{E_j + M}}{2\lambda_j \frac{E_j}{|E_j|}\sqrt{E_j-M}}C_j^{\mathrm{down}}.
\eeqn
Consequently, the spinor $u_j^\lambda$ with the helicity $\lambda$ can be written as follows
\beqn
u_j^\lambda (\rho,\varphi) = C_j
	\begin{pmatrix}
	\bE_+ \phi_j^\lambda \\
	2\lambda_j \frac{E_j}{|E_j|} \bE_- \phi_j^\lambda
	\end{pmatrix}
	\label{eq:u:j}
\eeqn
with $\bE_\pm = \sqrt{1\pm \frac{M}{E}}$ and an overall constant $C_j$, which is determined by an orthogonal condition. Notice that the prefactor $\alpha _j$ in Eq.(\ref{eq:phi:j}) is absorbed into $C_j$.

The spinor $u_j$ which satisfies the MIT boundary condition~\eq{eq:MIT:bc} can be constructed in terms of the linear combination
\beqn
u_j(\rho,\varphi) = b^+_j u^+_j(\rho,\varphi)  + b^-_j u^-_j(\rho,\varphi)\,.
\label{eq:linear}
\eeqn

Substituting the eigenmode~\eq{eq:U} and \eq{eq:linear} into the boundary condition~\eq{eq:MIT:bc} as $\psi \equiv U_j$ and using the explicit form of the eigenspinors~\eq{eq:u:j} we get a matrix equation for the coefficients $b^\pm$ with the solution~\eq{eq:linear}:
\beqn
& & 
\bE_+ \left(b^+_j \phi^+_j + b^-_j \phi^-_j\right){\biggl |}_{\rho = R} \nonumber \\
& = & -\frac{iE}{|E|}\bE_- \left(b^+_j \sigma^{\rho}  \phi^+_j -  b^-_j \sigma^{\rho} \phi^-_j \right){\Biggl |}_{\rho = R}
\label{eq:BC:explicit}
\eeqn
where $\sigma^{\rho}$ is given in Eq.~\eq{eq:sigma:rho}.
The matrix equation~\eq{eq:BC:explicit} can also be represented in the form:
\begin{widetext}
\beqn
\begin{pmatrix}
 i \frac{E}{|E|} \bE_-e^{-i\varphi}f_{j +}^+\mM^+{+}
 \bE_+f_{j-}^+\mM^- 
&  -i \frac{E}{|E|} \bE_-e^{-i\varphi}f_{j+}^-\mM^+ {+} 
\bE_+f_{j-}^-\mM^- \\[3mm]
  i \frac{E}{|E|} \bE_-e^{i\varphi}f_{j-}^+\mM^- {+} 
\bE_+f^+_{j+}\mM^+
& - i \frac{E}{|E|} \bE_-e^{i\varphi}f_{j-}^-\mM^- {+} 
\bE_+f^-_{j+}\mM^+
\
\end{pmatrix}
\!\!
\begin{pmatrix}
\\[-3mm]
b^+ \\[3mm]
b^-\\[1mm]
\end{pmatrix}
{\Biggl |}_{\rho = R}{=} 0.
\qquad
\label{eq:BC:explicit:2}
\eeqn

We find that Eq.~\eq{eq:BC:explicit:2} has a nontrivial solution for $b^{\pm}$ if the quantity 
\beqn
q^B = \sqrt{E^2-M^2-k^2}R
\eeqn
satisfies the following relation:
\beqn
 \left\{ \begin{array} {ll}
(q^B)^2(\mM^+_{R})^2 \,-\, 4
(m+1) MR \mM^-_{R} \mM^+_{R} - 4(m+1)^2(\mM^-_{R})^2 = 0, &\qquad m\ge 0 \,, \\[3mm]
(q^B)^2(\mM^-_{R})^2 - 4
m MR \mM^-_{R} \mM^+_{R} - 4 m^2(\mM^+_{R})^2 = 0, &\qquad m < 0 \,,
\end{array}
\right.
\label{eq:M}
\eeqn
\end{widetext}
where
\beqn
\begin{array}{rcl}
\mM^+_{R} \equiv \mM^+ {{\Bigl|}_{\rho = R} } & = & \mM\left(a_j^+, |m_j+1|+1, \phi_B/\phi_0 \right)\,, 
\\[3mm]
\mM^-_{R} \equiv \mM^- {{\Bigl|}_{\rho = R} } &= & \mM\left(a_j^-, |m_j|+1,  \phi_B/\phi_0 \right)\,.
\end{array}
\label{eq:M:plus:minus}
\eeqn
The magnetic field enters the spectrum in terms of the ratio  
\beqn
\frac{\phi_B}{\phi_0} \equiv \frac{eB R^2}{2}\,.
\label{eq:phi:B:0}
\eeqn
of the magnetic flux the crosssection of the cylinder
\beqn
\phi_B = \pi B R^2\,,
\label{eq:phi:B}
\eeqn
and the elementary magnetic flux
\beqn
\phi_0 = \frac{2 \pi}{e}\,,
\label{eq:phi:0}
\eeqn
(we remind that in our units $\hbar = 1$).

Since the dimensionless quantity $q^B$ is discretized in accordance with effects of both the boundary condition and the Landau quantization, it can be labeled by the angular momentum number $m$ and the root number $l=1,2,3, \dots$, i.e. $q^B_{ml}$. 

The zero solutions of Eq.\eq{eq:M},  $q^B_{ml} = 0$ are achieved at specific values of the fermion masses $M = M^{(m)}_c$ with
\beqn
 M_c^{(m)} =
 \left\{ \begin{array} {ll}
 -\frac{m+1}{R}\frac{1}{\mM \left(1, m+2, \phi_B/\phi_0\right)}, &\qquad m\ge 0 \,, \\[3mm]
  \frac{m}{R}\frac{e^{\phi_B/\phi_0}}{\mM \left(-m, -m+1, \phi_B/\phi_0\right)}, &\qquad m < 0 \,,
\end{array}
\right.
\label{eq:zero:M}
\eeqn
where we used the properties $\mM (0,b,z) = 1$ and $\mM(a,a, z) = e^{z}.$ 
In the limit of vanishing magnetic field, $eB \rightarrow 0,$ we can recover the result~\eq{eq:M:c} for $M^{(m)}_c$ using the property $ \mM(a,b,z) = 1 + O(z)$ valid for $z \to 0$. In the limit of strong magnetic field, $eB \rightarrow \infty$, the mass becomes 
\beqn
M_c^{(m)} =
 \left\{ \begin{array} {ll}
 -\frac{e^{-\phi_B/\phi_0}\left(\phi_B/\phi_0\right)^{m+1}}{Rm!} \rightarrow 0, &\qquad m\ge 0 \,, \\[3mm]
 - \frac{1}{R} \frac{\phi_B}{\phi_0} \rightarrow -\infty, &\qquad m < 0 \,, \qquad
\end{array}
\right.
\label{eq:lim:zero:M}
\eeqn
where we used the asymptotic expansion $\mM(a,b,z) \sim (\Gamma (b)/\Gamma(a)) e^z z^{a-b}$ valid at $z \rightarrow \infty$ for all values of $a$ except for non-positive integer $a$.

We can recover Eq.~\eq{eq:J} from Eqs.~\eq{eq:M} and \eq{eq:M:plus:minus} in the limit of vanishing magnetic field $eB \to 0$ using the relations (valid for $q_B \neq 0$ and $n \geq 0$)
\beqn
& &  a^{\pm}_j  \xrightarrow{eB \to 0} -\frac{q_{B,j}^2}{2eBR^2},
\label{eq:limit:1} \\
& & \lim_{x \to 0} {}_1 F_1 \left(-\frac{y^2}{2 x},n+1; \frac{x}{2}\right) = n! \left( \frac{2}{y} \right)^n J_n(y), 
\label{eq:limit:2}
\eeqn
and $J_{-m}(x) = (-1)^m J_m(x)$.

The masses of the bulk and the edge states are given by the same formulae~\eq{eq:M:bulk} and, respectively, \eq{eq:M:edge} as in the case of the $B=0$ states (with the obvious change $q_{ml} \to q_{ml}^B$).  The quantity $\nu_m^B$ for the edge states in the background of magnetic field is defined similarly to the $B=0$ definition in Eq.~\eq{eq:q:nu:edge}:
\beqn
q^B_m = i \nu_m^{B}.
\label{eq:q:nu:edge:B}
\eeqn 

\subsection{Properties of the solutions}

In order to obtain the spectrum of free fermions in the cylinder in the presence of external magnetic field we solve Eqs.~\eq{eq:M} and \eq{eq:M:plus:minus} numerically.

\begin{figure}[!thb]
\hskip 1mm \includegraphics[scale=0.5,clip=true]{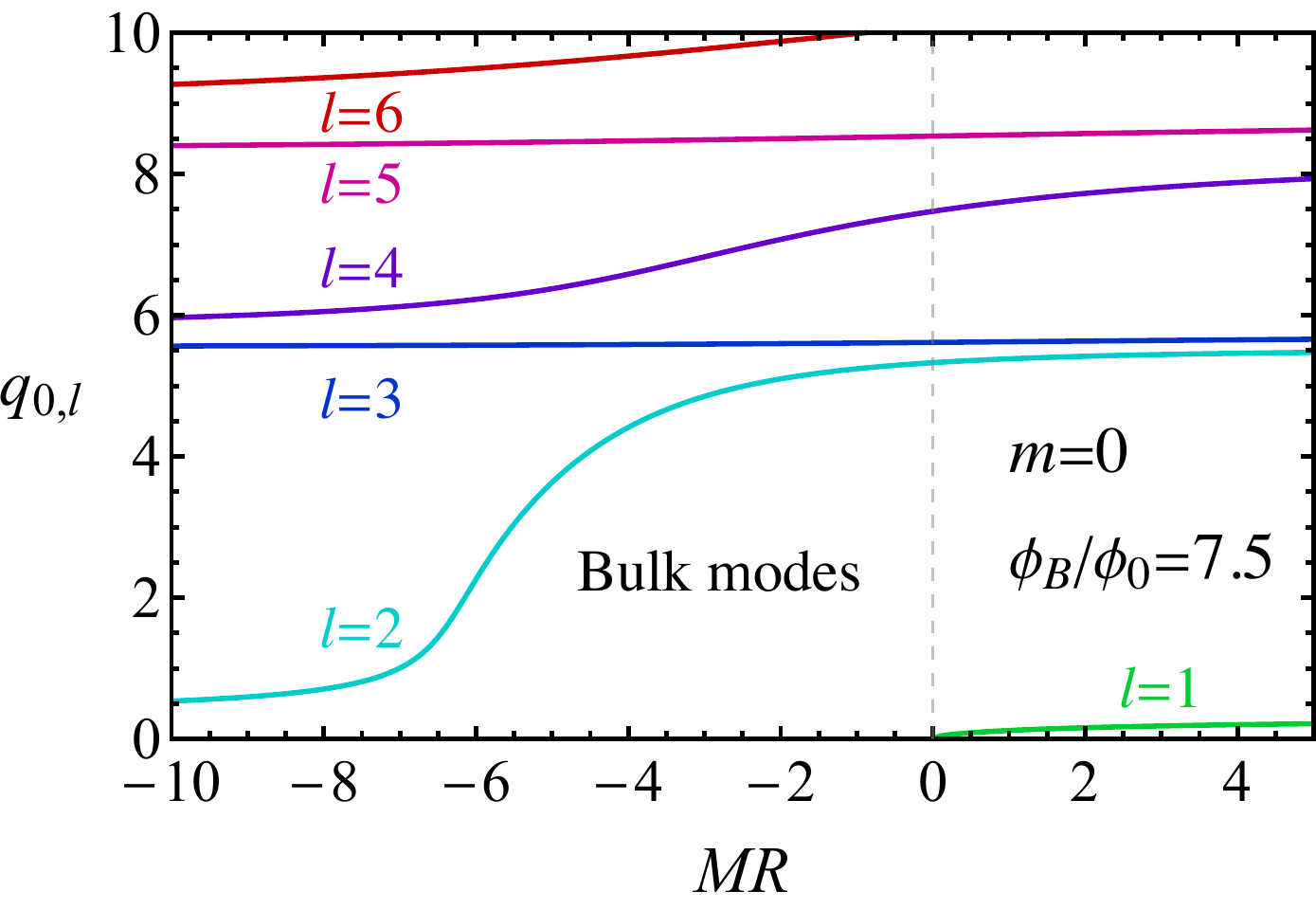} \\
\hskip 7mm (a) \\[3mm]
\includegraphics[scale=0.5,clip=true]{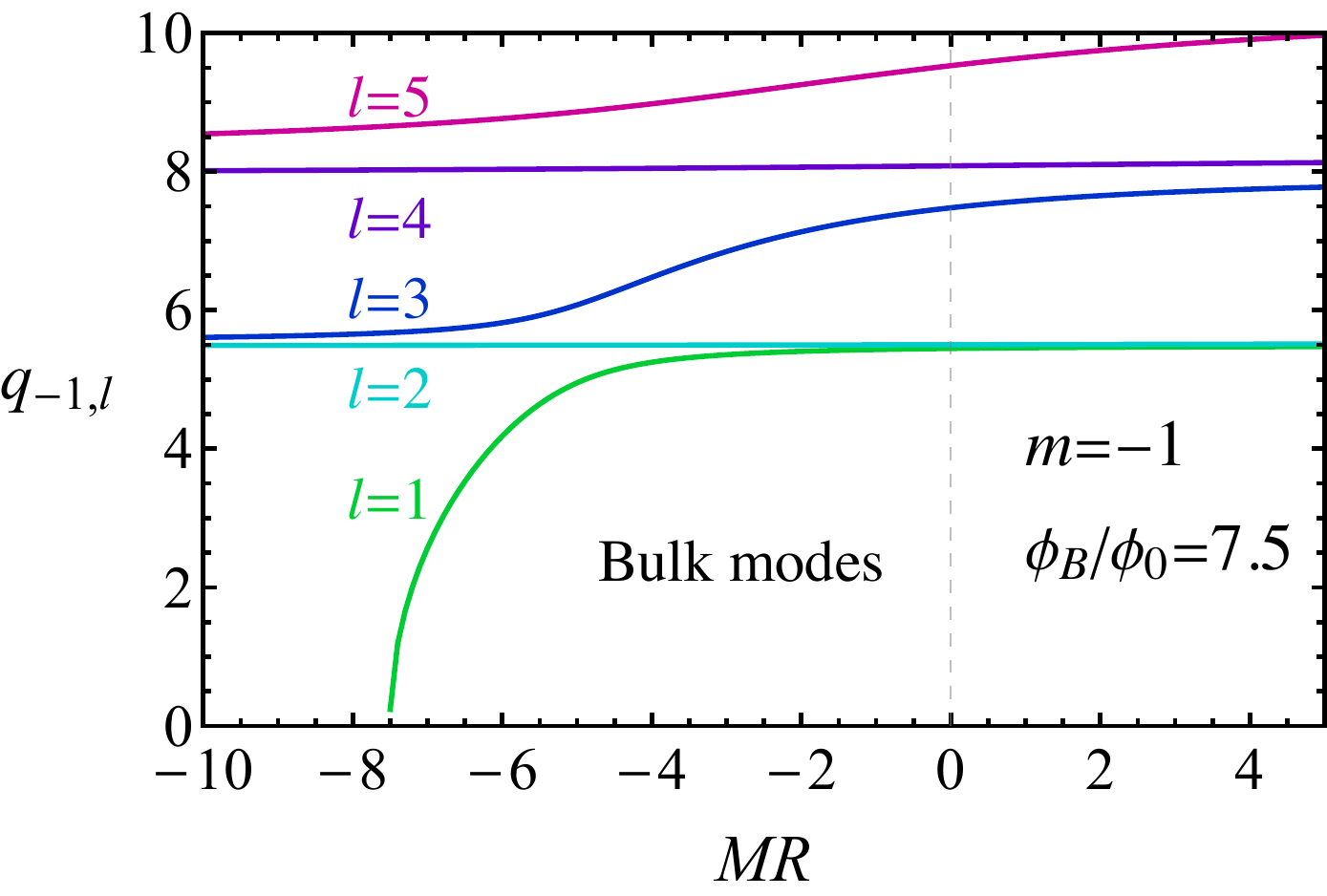}\\
\hskip 7mm (b)
\caption{The bulk $q^B_{ml}$ solutions of Eqs.~\eq{eq:M} and \eq{eq:M:plus:minus} vs. the fermion mass $M$ in the background of magnetic flux~\eq{eq:phi:B:0} $\phi_B = 7.5 \phi_0$ for (a) $m=0$ and (b) $m=-1$ orbital numbers and various radial excitation numbers $l$.}
\label{fig:q:B15}
\end{figure}

\begin{figure}[!thb]
\hskip 2mm \includegraphics[scale=0.5,clip=true]{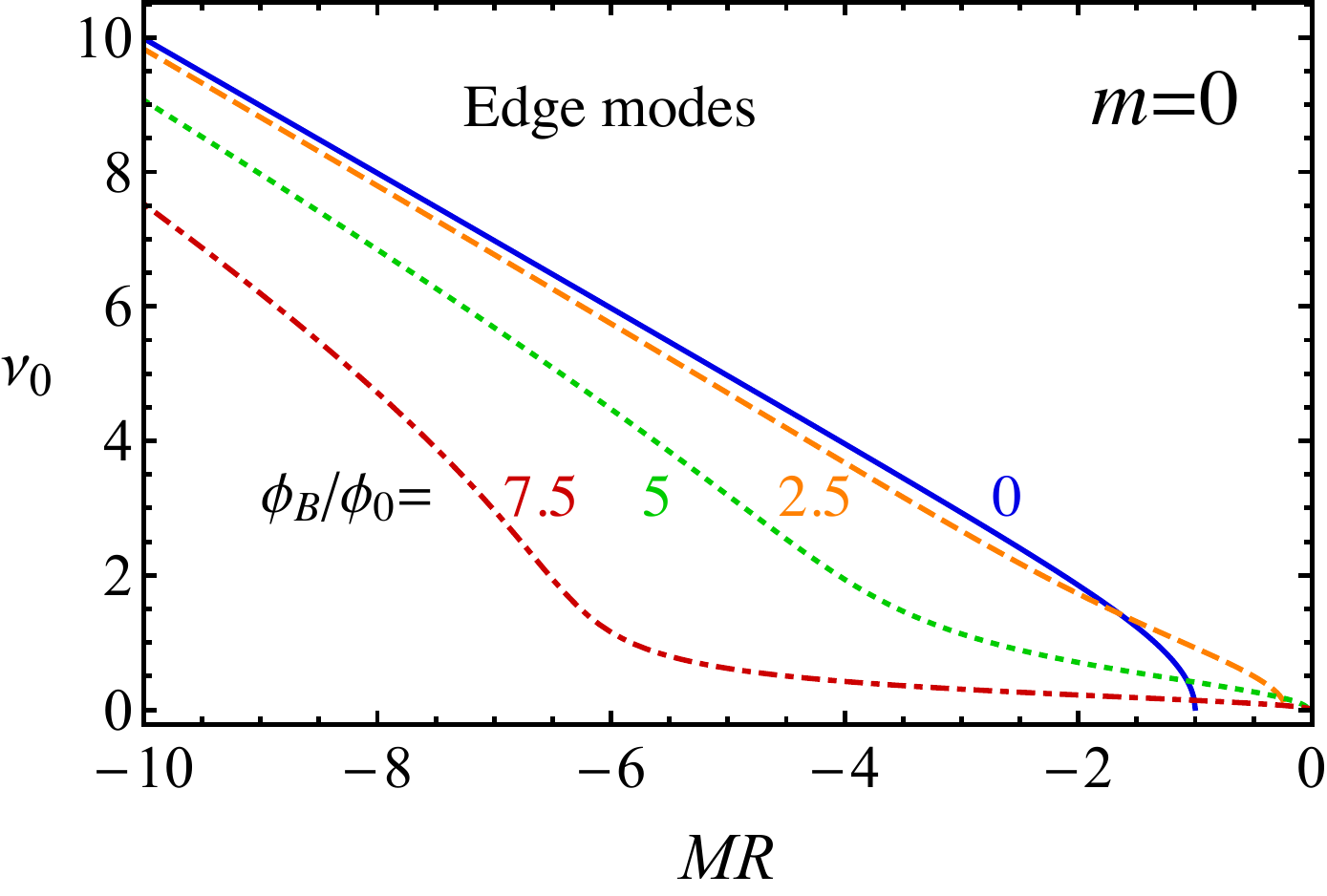} \\
\hskip 7mm (a) \\[3mm]
\includegraphics[scale=0.5,clip=true]{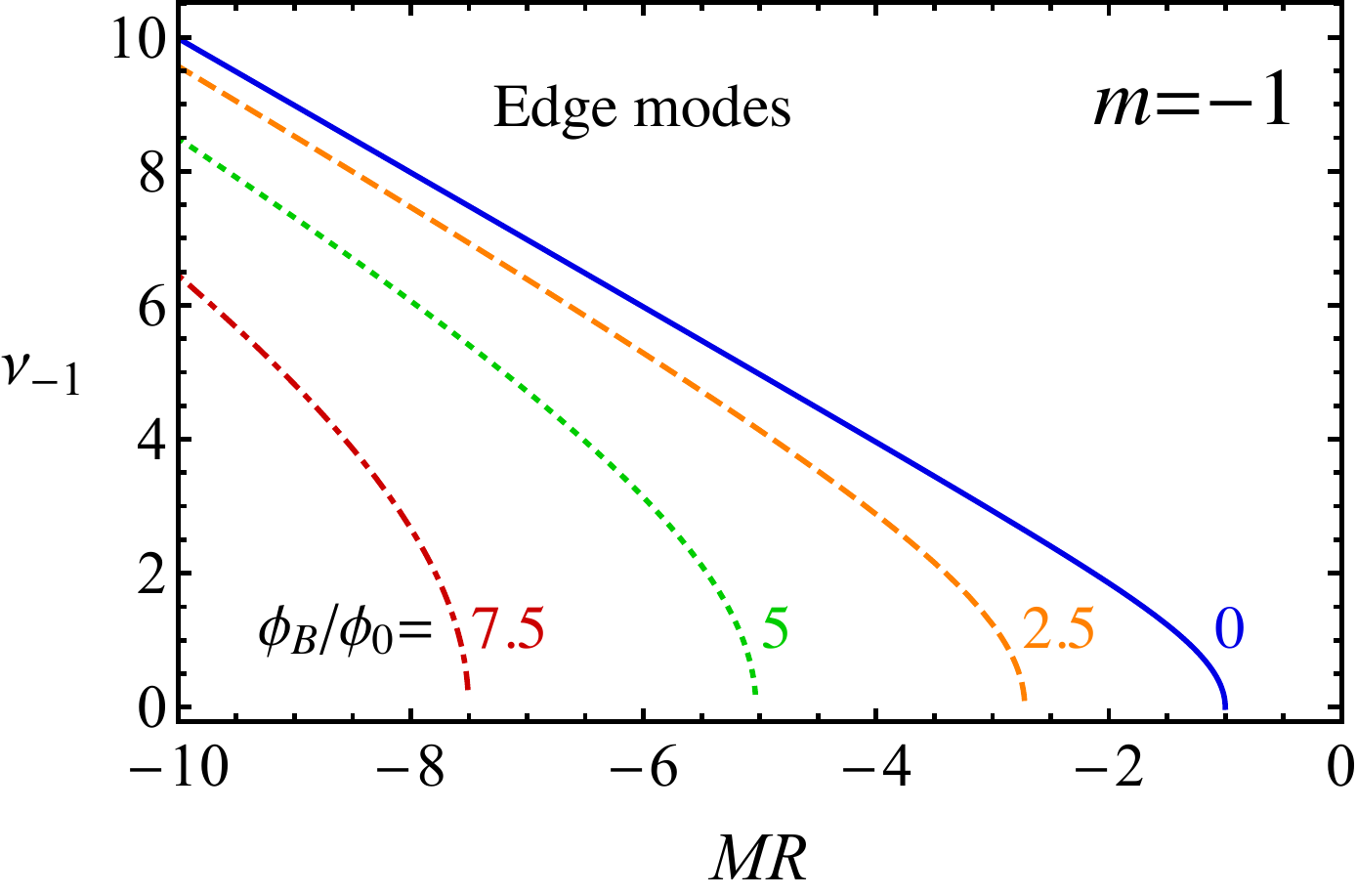} \\
\hskip 7mm (b)
\caption{The edge $\nu^B_{m}$ solutions~\eq{eq:q:nu:edge:B} of Eqs.~\eq{eq:M} and \eq{eq:M:plus:minus} vs. the fermion mass $M$ in the background of different magnetic fluxes~\eq{eq:phi:B:0} $\phi_B$ for (a) $m=0$ and (b) $m=-1$ orbital numbers.}
\label{fig:nu:magnetic}
\end{figure}

In Figs.~\ref{fig:q:B15} and \ref{fig:nu:magnetic} we show the behavior of, respectively, the bulk solutions $q^B_{ml}$ and the edge solutions $\nu_m^B$ for the orbital angular momentum $m=0$ (which represents the qualitative behavior of all $\mu_m>0$ modes) and $m=-1$ (which characterizes general properties of the solutions with $\mu_m < 0$) at nonzero magnetic field. These quantities at zero magnetic field were shown in Fig.~\ref{fig:qmls}.

We notice the following effects of background magnetic field on the bulk modes:
\begin{enumerate}
\item[(i)] {\sl Critical mass:} at zero magnetic field the ground states ($l=1$) disappear at the quantized critical masses $M_c$ given in Eq.~\eq{eq:M:c}. As the magnetic field becomes stronger the critical masses $M_c$ deviate from their $B=0$ values: for $eB>0$ the critical masses for the modes with a positive angular momentum $\mu_m>0$ tend to zero, $M_c \to 0$, while the critical masses of the $\mu_m < 0$ modes tend to a negative infinity, $M_c \to -(\phi_B/\phi_0)/R$. The behaviors are consistent with the analytical results given in Eq.\eq{eq:lim:zero:M}. One can show that at $eB < 0$ the modes with $\mu_m>0$ and $\mu_m<0$ swap their places as $M_c \to - \infty$ for the former and $M_c \to 0$ for the later. 

\item[(ii)] {\sl Level degeneracy:} at large positive or negative values of the fermion mass, $M \to \pm \infty$, the levels are grouping into pairs. This is a natural consequence of growing mass of the bulk levels~\eq{eq:M:bulk}. As the mass become large, the bulk states become more localized in space and they become less sensitive to the presence of the boundary of the cylinder. Then the energy spectrum shares a natural similarity with the Landau levels in a boundless space where the spin-up and spin-down states of the excited levels are double-degenerate in energy. 

\end{enumerate}

The behavior of the edge modes $\nu_m^B$ at values of magnetic field -- or, equivalently, the magnetic flux $\phi_B$, Eq.~\eq{eq:phi:B} -- is shown in Fig.~\ref{fig:nu:magnetic}. The mentioned properties of the critical mass is well consistent with the ones for the bulk modes, as expected. As the fermion mass $M$ decreases the quantities $\nu_m$ become linear functions of the mass $M$.

\begin{figure*}[!thb]
\begin{tabular}{cc}
\includegraphics[scale=0.525,clip=true]{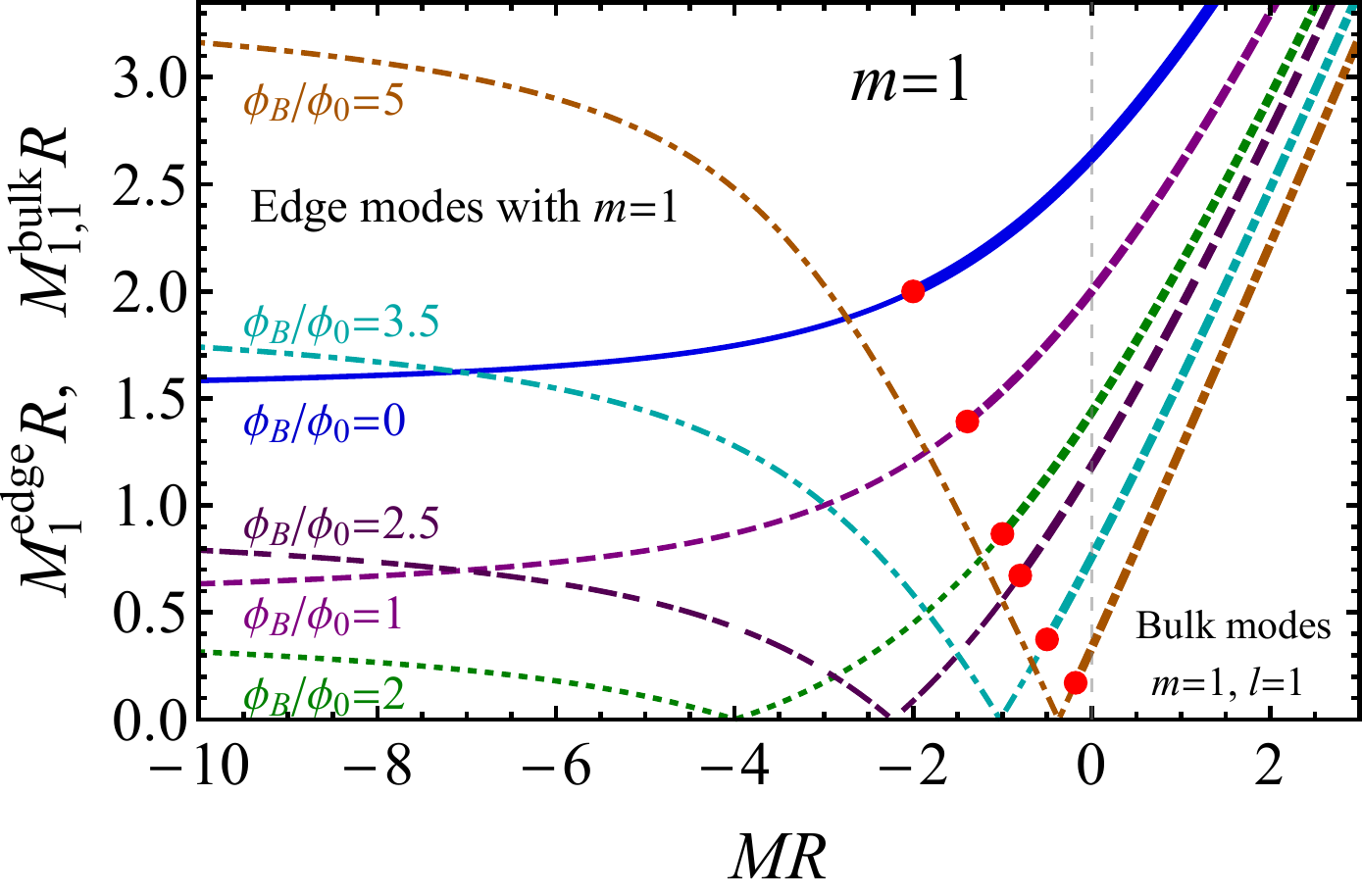} &
\hskip 10mm \includegraphics[scale=0.525,clip=true]{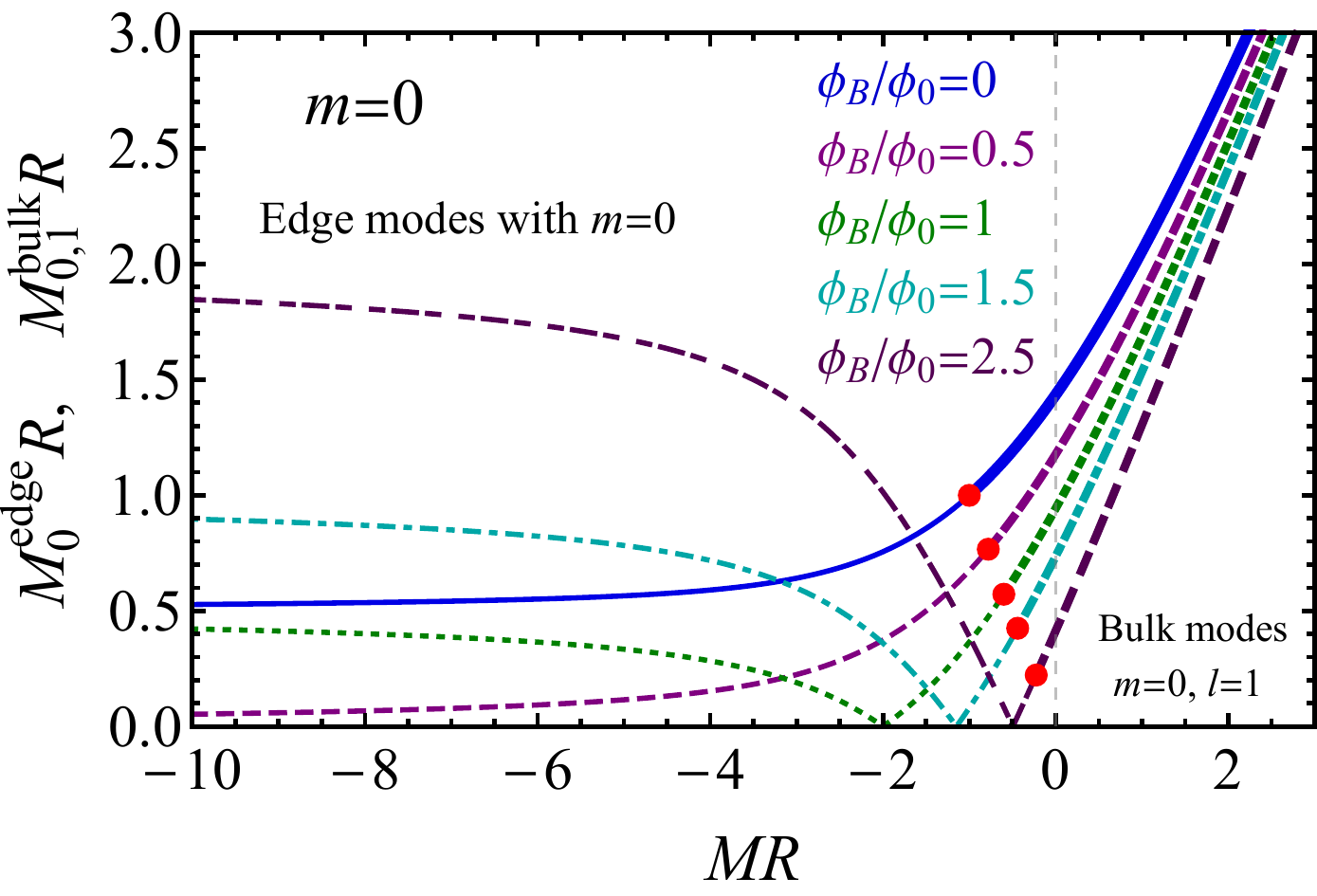} \\
\hskip 9mm (a) & \hskip 19mm (b) \\[3mm]
\includegraphics[scale=0.525,clip=true]{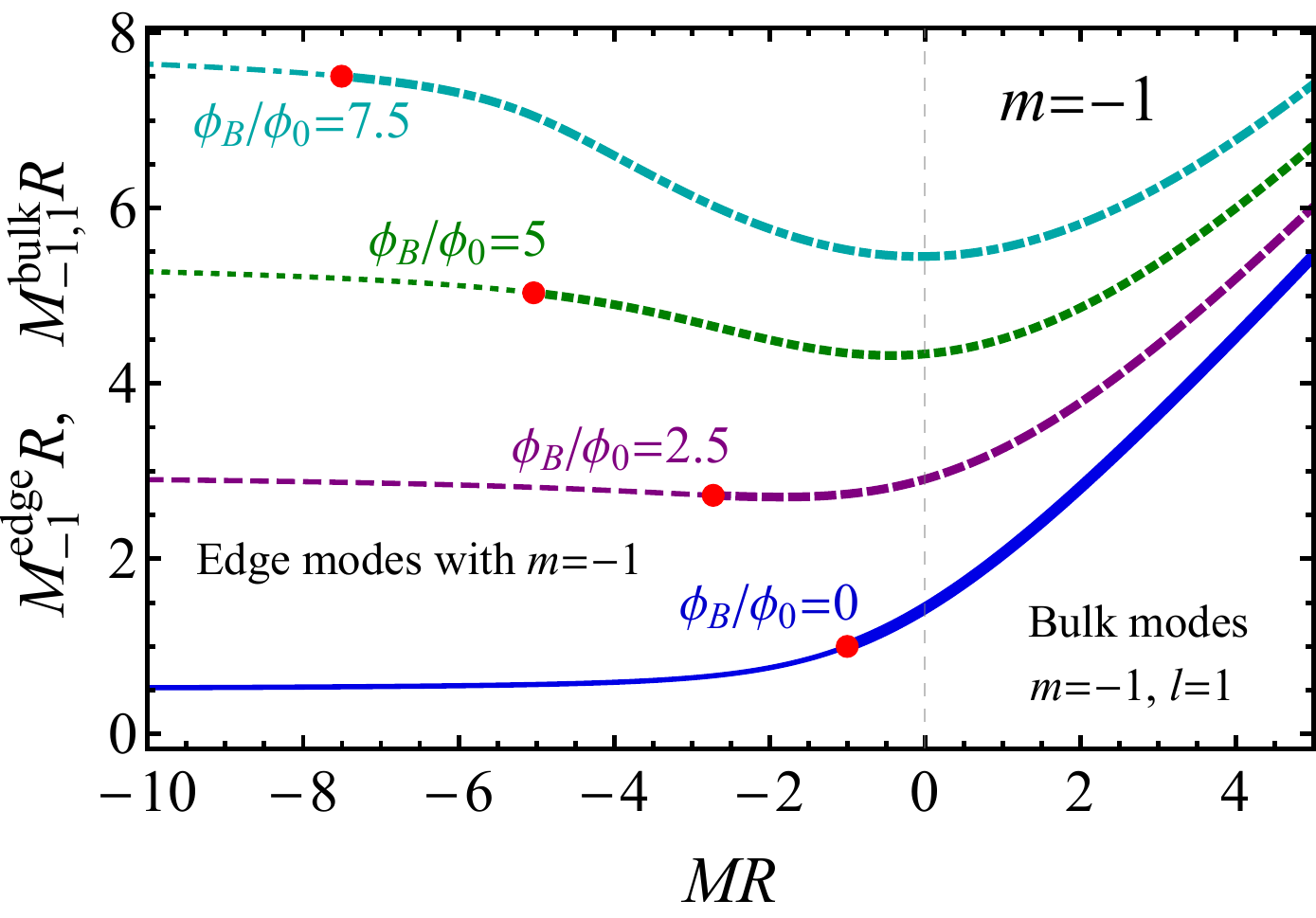} &
\hskip 10mm \includegraphics[scale=0.525,clip=true]{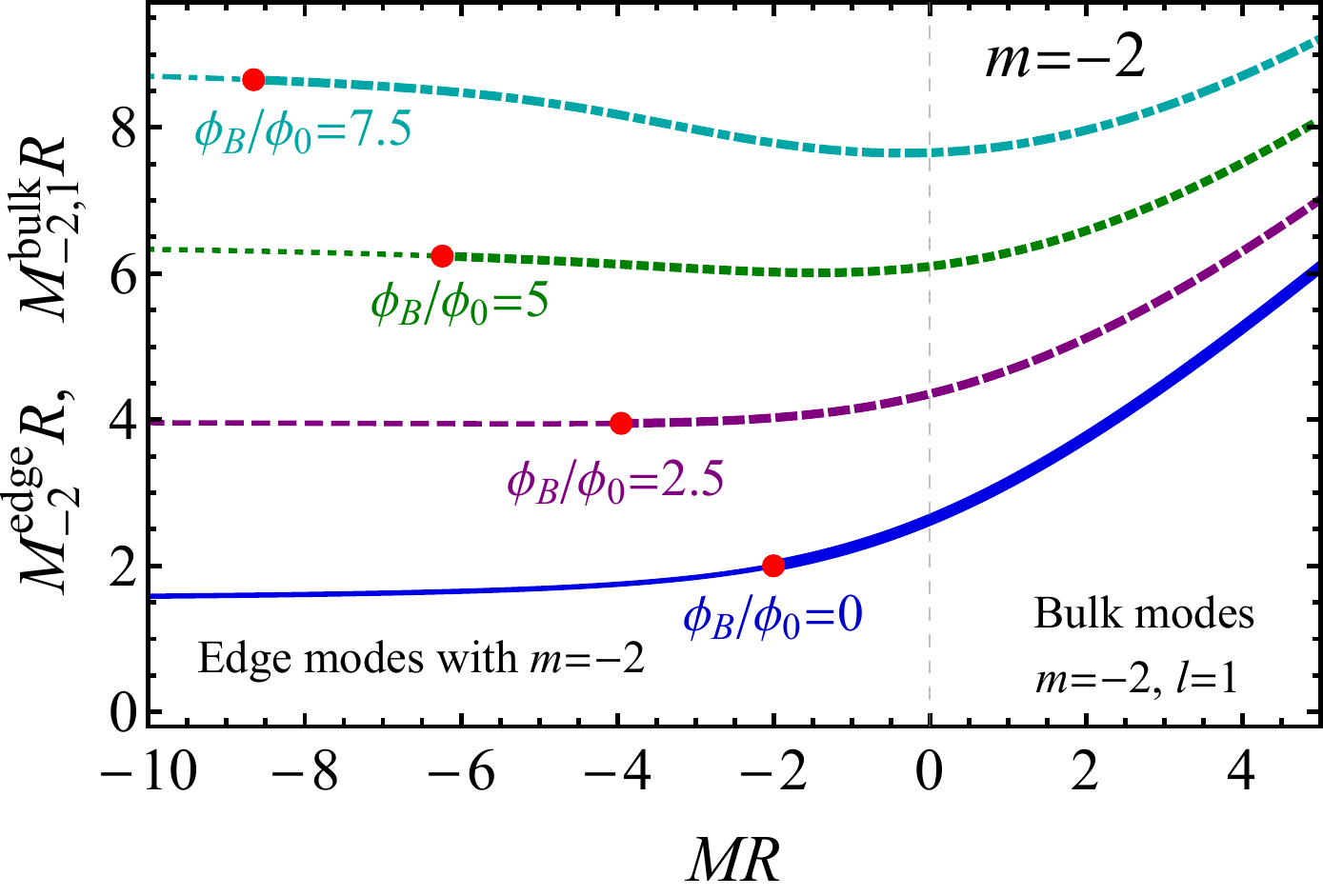} \\
\hskip 7mm (c) & \hskip 18mm (d) 
\end{tabular}
\caption{The masses of the lowest bulk ($l=1$) and edge states vs. the mass of the fermion $M$ for various values orbital angular momenta $m$ and magnetic field $B$. The bulk (edge) modes are shown by the thicker (thinner) lines while the positions where the bulk modes are converted to the corresponding edge modes are marked by the red points.}
\label{fig:masses:magnetic}
\end{figure*}

In Fig.~\ref{fig:masses:magnetic} we show the masses of the lowest ($l=1$) bulk modes~\eq{eq:M:bulk} and the edge modes~\eq{eq:M:edge} as the functions of the fermion mass $M$ at various values of magnetic field $B$. We notice the following remarkable properties of these quantities:

\begin{figure}[!thb]
\includegraphics[scale=0.525,clip=true]{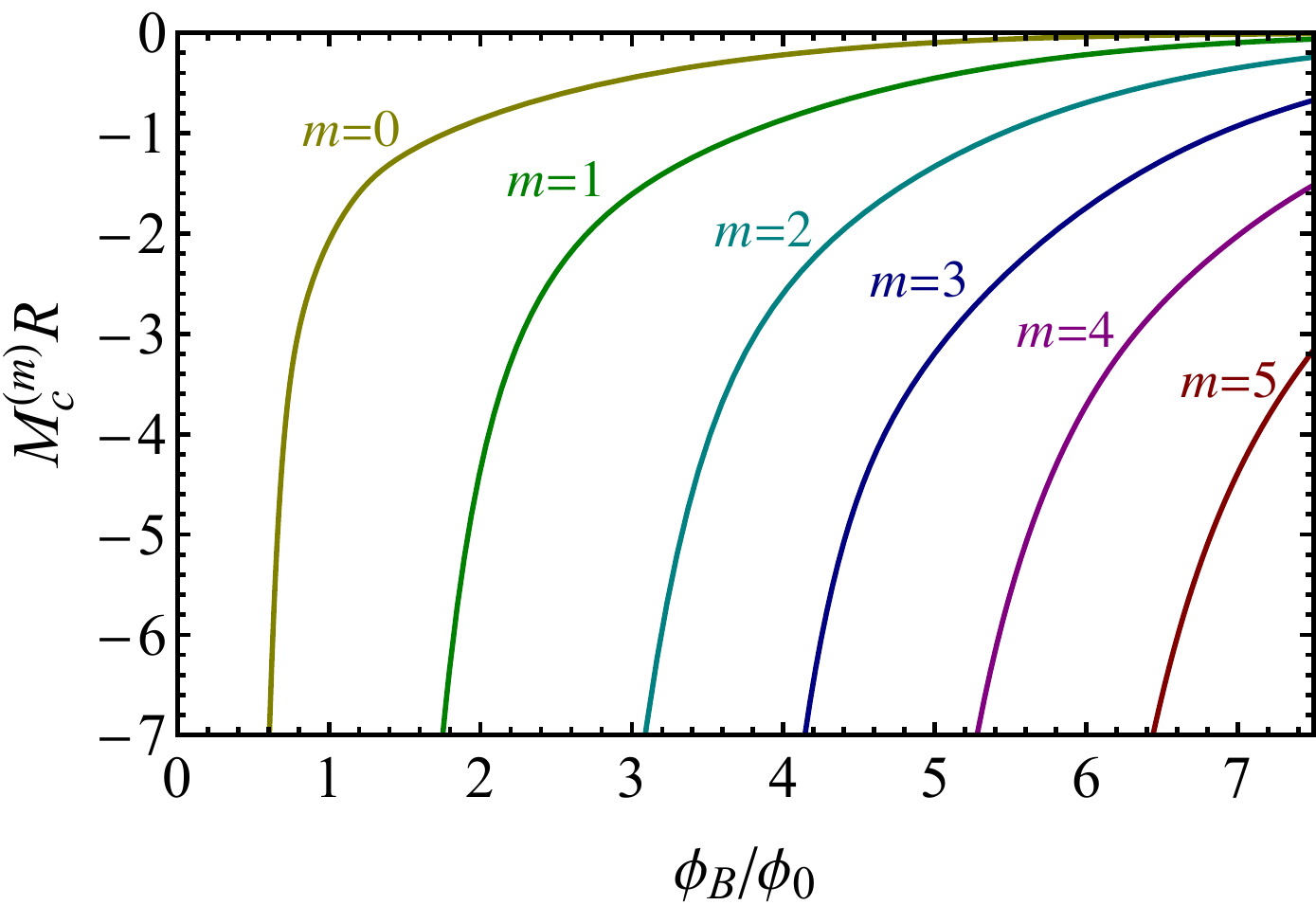}
\caption{The values of the fermion masses $M=M_c^{(m)}$ at which the masses of the edge modes vanish~\eq{eq:M:edge:vanish} vs the magnetic flux~$\phi_B$ for various values of orbital momentum $m$.}
\label{fig:M:crit}
\end{figure}

\begin{itemize}

\item[(i)] Masses for the modes with negative angular momenta $\mu_m$ (i.e. with $m=-1,-2, \dots$) behave regularly as the $l=1$ bulk modes are transformed into the edge modes at certain critical masses $M = M_c^{(m)}(B)$. These critical masses are growing in absolute value (and negatively-valued) functions of magnetic field. At large enough strengths of the background magnetic field the masses of the bulk modes experience, as functions of the fermion mass $M$, a global minimum.

\item[(ii)] At positive values of the angular momenta $\mu_m$ (i.e. at $m=0,1,2, \dots$) the masses of the edge modes behave rather irregularly. In particular, they vanish at certain mass $M = M_{c}^{(m)}(B)$,
\beqn
M^\ed_m\left(M_c^{(m)}(B)\right) = 0\,, \qquad m\geqslant 0\,.
\label{eq:M:edge:vanish}
\eeqn
In Fig.~\ref{fig:M:crit} we plot,  for a few values of $m$, the masses of fermions $M=M_{c}^{(m)}(B)$ at which the mass of the edge mode become zero (effectively, the massive edge mode becomes the zero mode). These masses are growing (in absolute value) negative-valued functions of the magnetic field $B$. 

\item[(iii)] At large negative values of the fermion mass, $M \to - \infty$, the masses of the edge states become qualitatively independent on the fermionic mass $M$.

\end{itemize}
Notice that all these properties are valid for positive magnetic field $eB>0$. For the negative magnetic field, $eB<0$, the modes with positive and negative magnetic momenta $\mu_m$ swap their places. 

As in the absence of magnetic field, in the limit of a large (negative) fermionic mass $M \to -\infty$ the masses of the edge modes remain finite contrary to the excited $l\geqslant 2$ bulk modes which become massive~\eq{eq:M:bulk} and decouple from the system. Moreover one can check numerically that in this limit the mass spectrum of the edge states fits a simple analytical function:
\beqn
M^\ed_{\infty,m}(B) = \lim_{M \to - \infty} M^\ed_m(B) =  \left|\mu_m - \frac{\phi_B}{\phi_0} \right|\frac{1}{R}. \qquad
\label{eq:M:edge:infty:B}
\eeqn
In fact, we can obtain the result~\eq{eq:M:edge:infty:B} analytically by using the large $a$ expansion of $\mM (a,b,z)$~\cite{ref:Temme}:
\beqn
&{\mM}&\left(a,b,z\right) \notag \\
&=&\left(z/a\right)^{(1-b)/2}\frac{e^{z/2}%
\Gamma\left(1+a-b\right)\Gamma\left(b\right)}{\Gamma\left(a\right)}\*\Biggl[I_{b-1}\left(2\sqrt{az}%
\right) \notag \\ 
&-&\sqrt{\frac{z}{a}}I_{b}\left(2\sqrt{%
az}\right)\left(\frac{b}{2}-\frac{z}{12}\right) + O(a^{-1})\Biggr] 
 \label{eq:Ma:expansion}
\\
&=&\left(z/a\right)^{(1-b)/2}\frac{e^{z/2}%
\Gamma\left(1+a-b\right)\Gamma\left(b\right)}{\Gamma\left(a\right)}\*\Biggl[\frac{e^{2\sqrt{az}}}{2\sqrt{\pi\sqrt{az}}} \notag \\ 
&\times&\left(1+\frac{1-4(b-1)^2}{16\sqrt{az}} -\sqrt{\frac{z}{a}}\left(\frac{b}{2}-\frac{z}{12}\right) + O(a^{-1})\right) \Biggr].\notag
\eeqn
Substituting Eqs.~\eq{eq:Ma:expansion} and \eq{eq:Im:expansion} into Eq.~\eq{eq:M}, we obtain the solution of $\nu_m$ in terms of the expansion of a large negative mass $M$:
\beqn
\nu^\ed_m = |M| R - \frac{(\mu_m - \frac{\phi_B}{\phi_0})^2}{2 |M| R} + O\Bigl(\bigl(M R\bigr)^{-2}\Bigr)\,, 
\label{eq:nu:edge:infty:M}
\eeqn
which leads to Eq.\eq{eq:M:edge:infty:B}.
  
\begin{figure}[!thb]
\includegraphics[scale=0.525,clip=true]{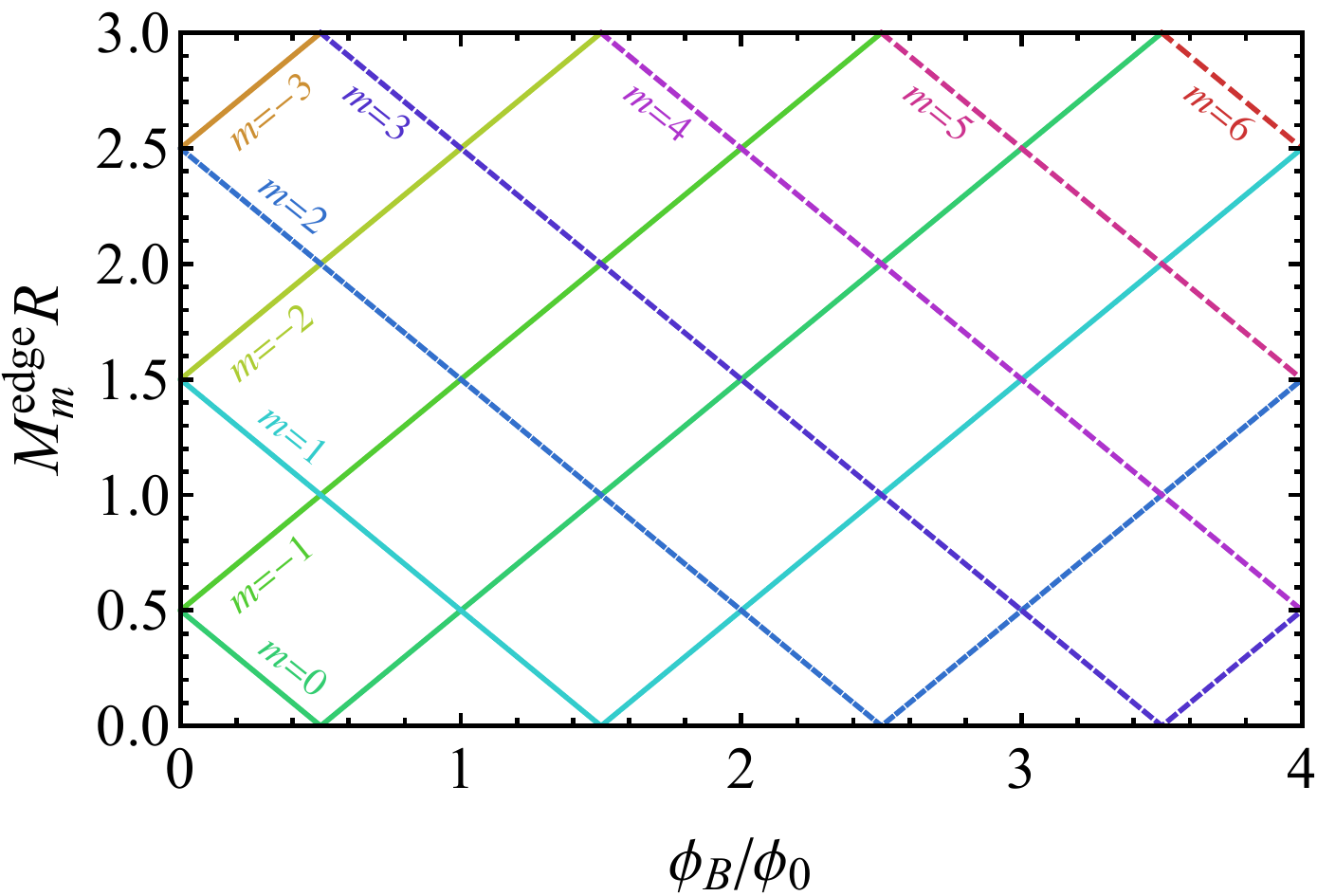}
\caption{The masses of the edge modes~\eq{eq:M:edge:infty} as functions of magnetic field $B$ in the limit $M \to - \infty$.}
\label{fig:edge:masses:infinite}
\end{figure}

The masses of the edge states depend on the angular magnetic moment $\mu_m$ of the mode and the Aharonov--Bohm phase $\vartheta = \phi_B/\phi_0$. In the limit of vanishing magnetic field, $\phi_B = 0 $, Eq.~\eq{eq:M:edge:infty:B} matches with the $B=0$ result~\eq{eq:M:edge:infty}. The mass spectrum of the edge states~\eq{eq:M:edge:infty:B} in the $M \to \infty$ limit is shown in Fig.~\ref{fig:edge:masses:infinite}.

\section{Edge modes and rotation }
\label{sec:edge:rotation}

\subsection{Zero magnetic field}

In the limit of infinite negative mass $M$ the thermodynamic and rotational properties of the system are determined only by the edge modes. Indeed, the masses of the edge modes remain finite~\eq{eq:M:edge:infty} while the masses of the bulk modes tend to infinity implying that the latter do not contribute to the dynamics of the system. In the absence of magnetic field the energy of the edge modes~\eq{eq:E:edge} is given by the following simple expression:\footnote{In this section we consider only the positively defined branch of the energy eigenmodes $E = + |E|$ which corresponds to the particle edge states~\eq{eq:E:edge} both for vanishing~\eq{eq:E:edge:M:inf} and nonvanishing~\eq{eq:E:edge:M:inf:B} magnetic field.}
\beqn
E^\ed_m(k_z) = \sqrt{k_z^2 + \frac{\mu_m^2}{R^2}}\,,
\label{eq:E:edge:M:inf}
\eeqn
where $\mu_m$ is the angular momentum of the edge mode~\eq{eq:mu:m} and $m\in \Z$.

The thermodynamic effects of the edge modes are determined by the thermodynamic potential defined in the corotating, as opposed to the laboratory, reference frame (the latter fact is stressed by the tilde sign in $\widetilde F$):
\beqn
& & {\widetilde F}^\ed(\sigma;T,\Omega) = - \frac{T}{\pi R^2} \sum_{m \in \Z}  \int \frac{d k_z}{2 \pi}\, 
\label{eq:F:edge}\\
& & \hskip 12mm
\biggl[\ln \left( 1 + e^{- \frac{E^\ed_{m}(k_z) - \Omega \mu_m}{T}} \right) + 
(\Omega \to - \Omega)
\biggr]. \qquad
\nonumber 
\eeqn 
Below we omit the superscript ``edge'' in all our notations.

The angular momentum density is given by the derivative of the thermodynamic potential in the corotating reference frame~\cite{ref:LL5}:
\beqn
{\bs L} = - {\left( \frac{\partial {\tilde F}}{\partial {\bs \Omega}} \right)}_{T}\,.
\label{eq:angular:momentum}
\eeqn
Since the rotation axis ${\bs \Omega} = \Omega \, {\bs{e}}_z$ coincides with the symmetry axis of the cylinder ${\bs{e}}_z$, the angular momentum has only one nonzero component, ${\bs L} = (0,0,L_z)$. 

It is convenient to consider the density of the angular momentum per unit height of the cylinder:
\beqn
& & {\cal L}_z(\Omega) \equiv \pi R^2 L_z(\Omega) \label{eq:L:density} \\
& & \qquad\quad = \int_{-\infty}^\infty \frac{k_z}{2\pi} \sum_{m \in \Z} \mu_m \left[ f_{m,k_z} (\Omega,T)  - f_{m,k_z} (-\Omega,T)  \right],
\nonumber
\eeqn
where 
\beqn
f_{m,k_z} (\Omega,T)  = \frac{1}{e^{\frac{E_{m}(k_z) - \Omega \mu_m}{T}} + 1}\,,
\eeqn
is the occupation number of the fermionic edge mode.

The moment of inertia per unit height is related to the density of the angular momentum~\eq{eq:L:density} as follows:
\beqn
{\cal I}_z(\Omega) = \frac{{\cal L}_z(\Omega)}{\Omega}\,.
\label{eq:I:density}
\eeqn

\begin{figure}[!thb]
\includegraphics[scale=0.5,clip=true]{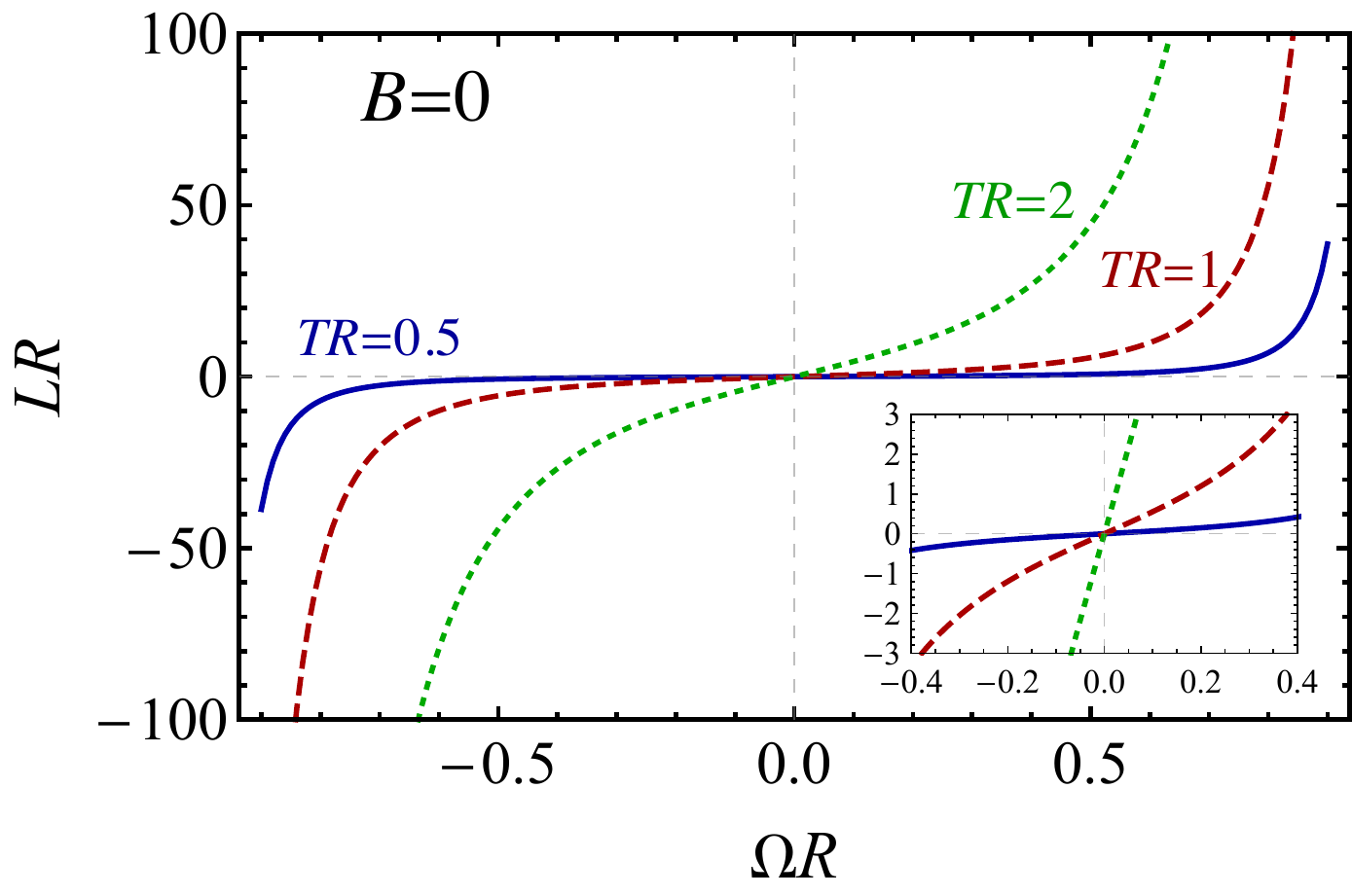} \\
\hskip 12mm (a)\\[7mm]
\hskip 7mm \includegraphics[scale=0.46,clip=true]{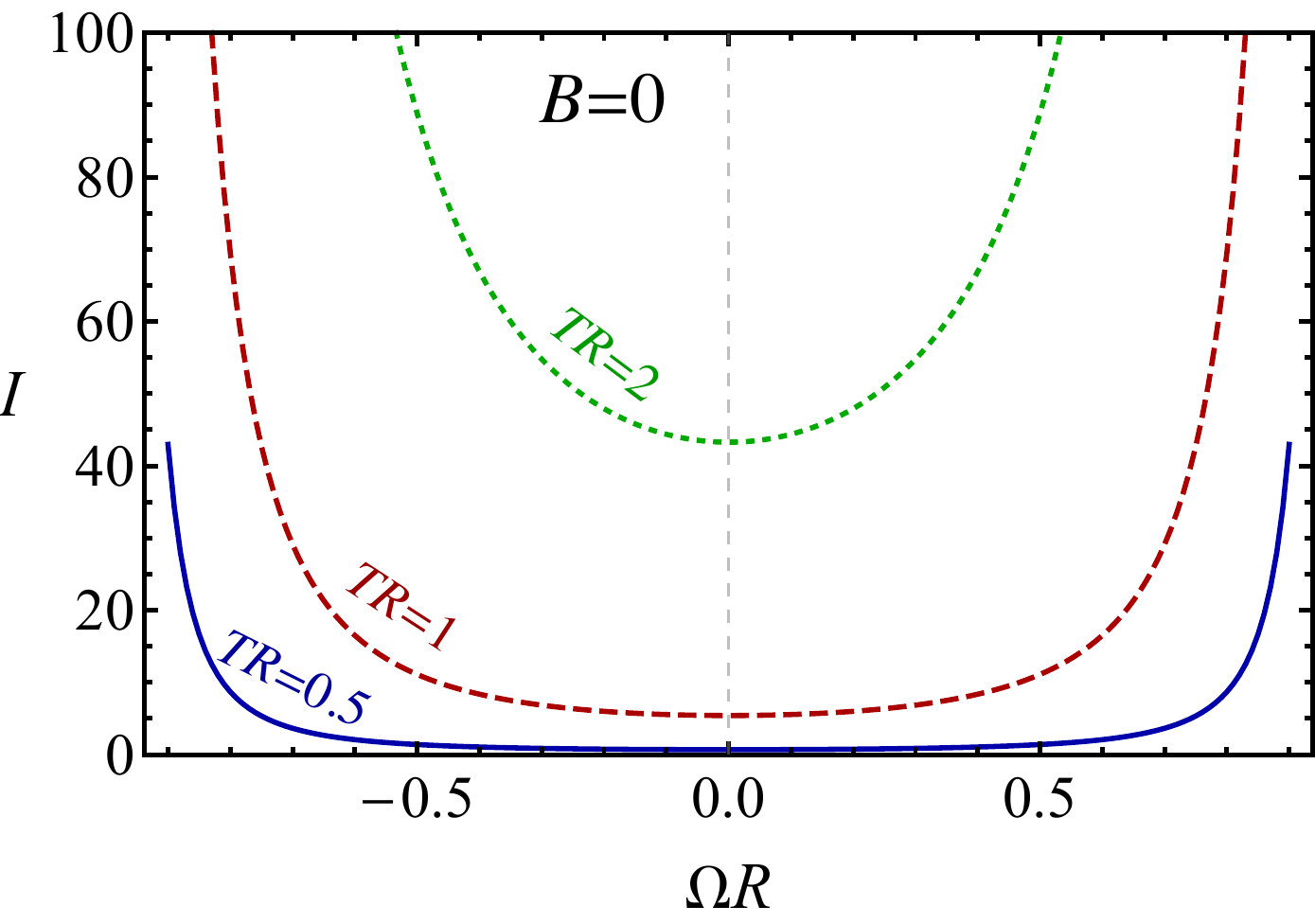} \\
\hskip 12mm (b)
\caption{Densities of (a) the angular momentum~\eq{eq:L:density} and (b) moment of inertia~\eq{eq:I:density} of the cylinder in the limit an infinite fermion mass $M \to \infty$ as the function of angular frequency $\Omega$ at various temperatures $T$ and zero magnetic field.}
\label{fig:angular:B0}
\end{figure}

The angular momentum~\eq{eq:L:density} and the moment of inertia~\eq{eq:I:density} at zero magnetic field are both shown in Fig.~\ref{fig:angular:B0}. These quantities are, respectively, odd and even functions with respect to the flips of the direction of rotation, $\Omega \to - \Omega$, because the thermodynamic potential~\eq{eq:F:edge} is an even function of $\Omega$.

\begin{figure}[!thb]
\includegraphics[scale=0.45,clip=true]{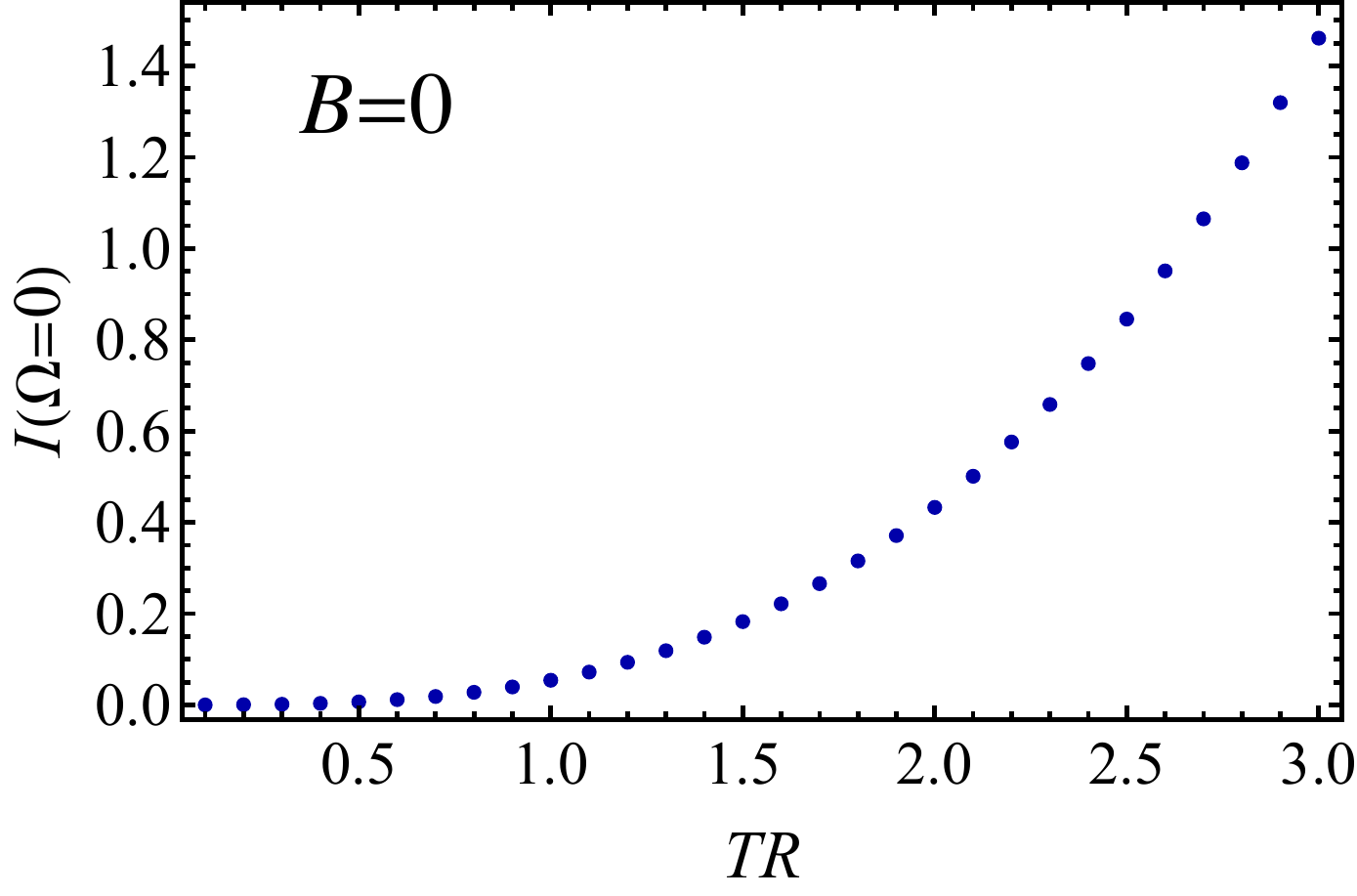}
\caption{Density of the moment of inertia~\eq{eq:I:density} at $\Omega =0$ vs. temperature $T$ at vanishing magnetic field $B=0$.}
\label{fig:moment:inertia:B0}
\end{figure}

In Fig.~\ref{fig:moment:inertia:B0} we show the density of the moment of inertia at zero angular momentum. The moment of inertia is a growing function of temperature because as temperature increases the heavier (energetic) modes may participate in rotation of the system.

\subsection{Effects of magnetic field}

In the presence of magnetic field the energy dispersion of the edge modes (in the limit of an infinite fermion mass $M \to \infty$) is given by the following formula:
\beqn
E^\ed_m(k_z) = \sqrt{k_z^2 + \frac{1}{R^2} {\left(\mu_m - \frac{\phi_B}{\phi_0} \right)}^2 }\,,
\label{eq:E:edge:M:inf:B}
\eeqn
where $\mu_m$ is the angular momentum of the edge mode~\eq{eq:mu:m} with $m\in \Z$.

\begin{figure}[!thb]
\includegraphics[scale=0.5,clip=true]{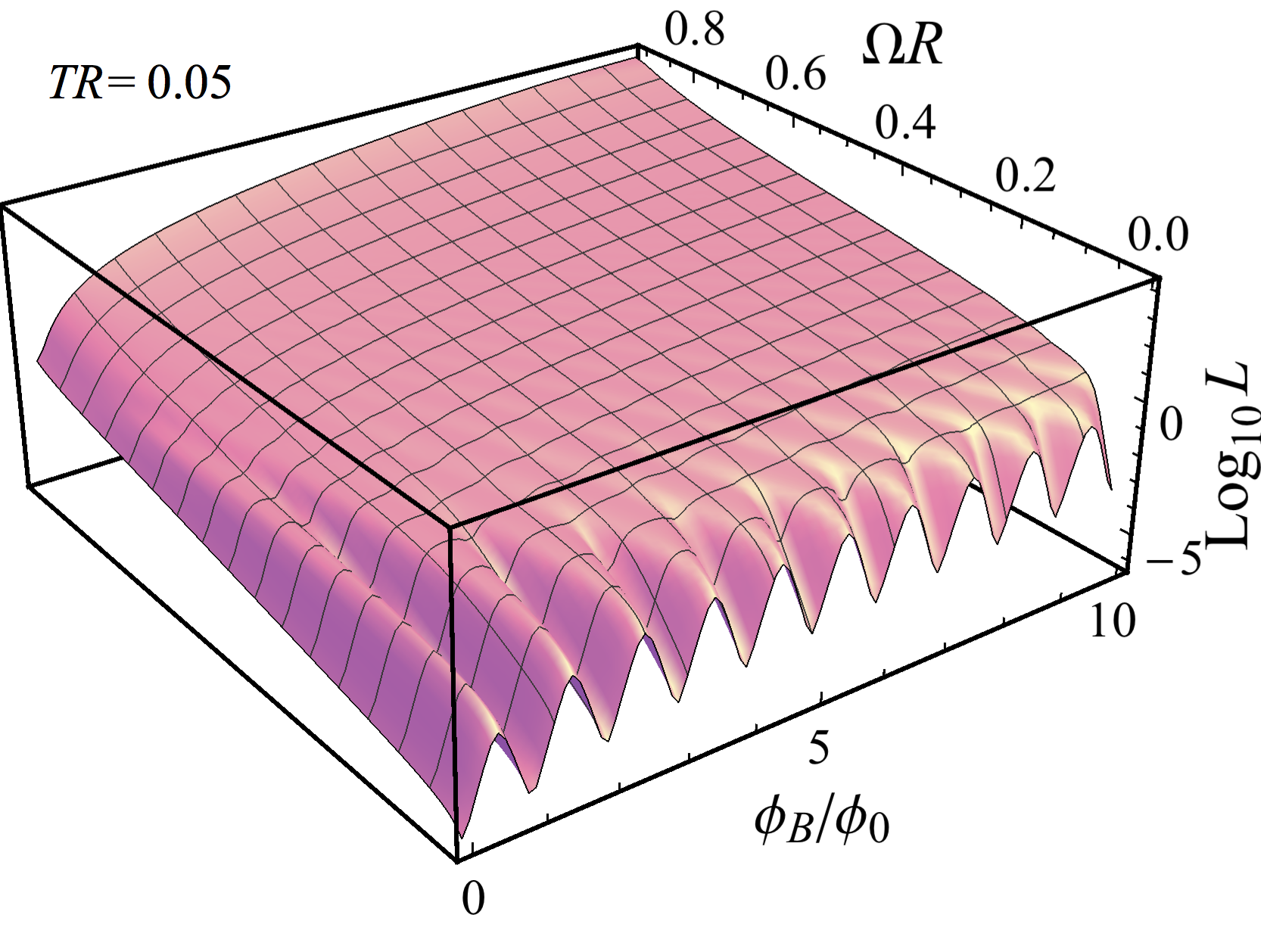}\\[-3mm]
(a)\\[5mm]
\includegraphics[scale=0.5,clip=true]{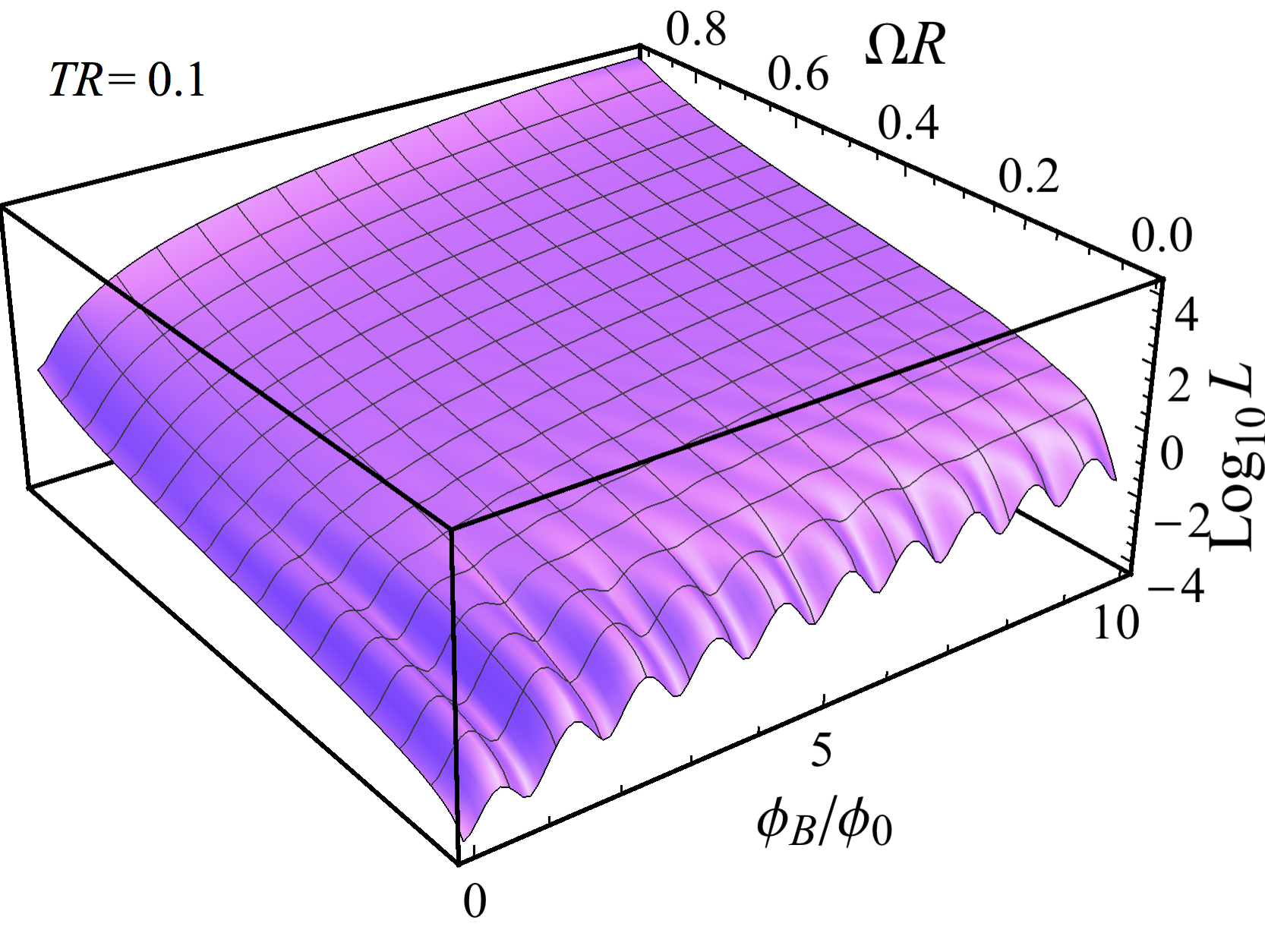}\\[-3mm]
(b)\\[5mm]
\includegraphics[scale=0.5,clip=true]{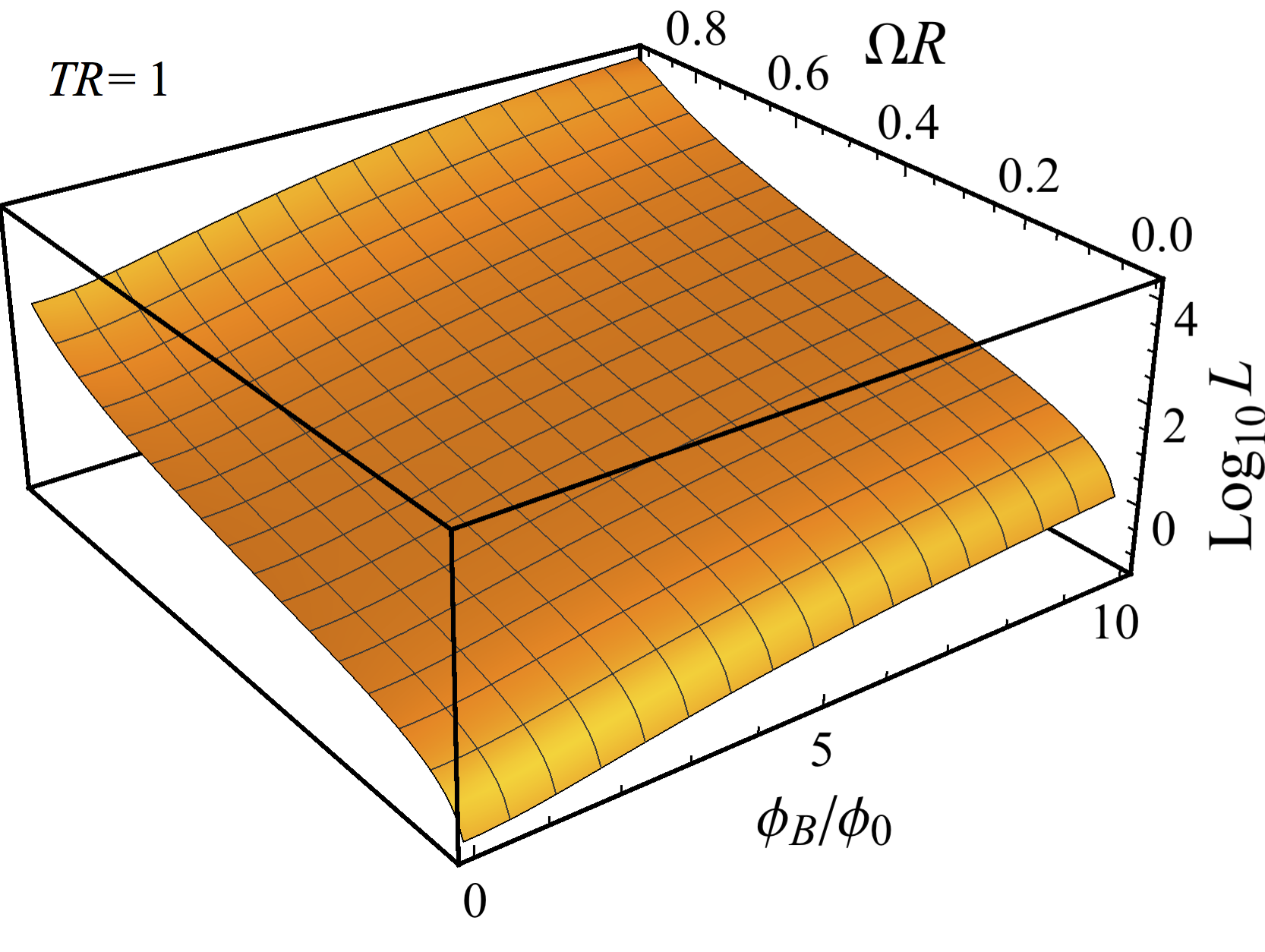}\\[-3mm]
(c)
\caption{Angular momentum $L$ of the edge modes per unit height of cylinder vs. angular frequency $\Omega$ and magnetic flux~$\phi_B$ at temperatures $TR = 0.05, 0.1, 1$ in the limit of infinite fermionic mass $M \to - \infty$ (the bulk modes are absent).}
\label{fig:angular:momentum:B:Omega}
\end{figure}

The angular momentum~\eq{eq:angular:momentum} can be readily calculated using the partition function~\eq{eq:F:edge} and dispersion~\eq{eq:E:edge:M:inf:B}. In Fig.~\ref{fig:angular:momentum:B:Omega} we show the angular momentum $L$ in the magnetic field - angular frequency $(B,\Omega)$ plane for temperatures $TR=0.05, 0.1, 1$. Naturally, the angular momentum is an increasing function of the angular frequency $\Omega$ for every fixed value of magnetic flux $\phi_B$ and for all temperatures $T$. 

At low temperatures $TR \lesssim 0.1$ and at slow rotations ($\Omega \sim 10^{-2}/R$) the angular momentum $L$ exhibits oscillating, but nonperiodic dependence on the value of magnetic flux, as it is clearly seen in Fig.~\ref{fig:angular:momentum:B:Omega}(a) and (b). The local minima and maxima of $L$ approximately correspond to the integer and, respectively, half-integer values of the ratio of magnetic flux $\phi_B$ and the elementary flux~\eq{eq:phi:B:0}. Apart from these oscillations, the value of $L$ slowly increases with strength of the background magnetic field. This quantum behavior is seen at sufficiently low temperatures: the lower temperature, the more pronounced oscillations. There is also certain small correlation between the magnetic field and the angular frequency seen in the range of middle frequencies, $\Omega R \sim 0.2$.

At higher temperatures $T R \sim 1$, shown in Fig.~\ref{fig:angular:momentum:B:Omega}(c), the magnetic-field induced oscillations of the angular momentum disappear completely.  At sufficiently fast rotations the oscillations disappear for all temperatures. In these cases the angular momentum is an increasing function of both magnetic field $B$ and angular frequency $\Omega$.

\begin{figure}[!thb]
\includegraphics[scale=0.55,clip=true]{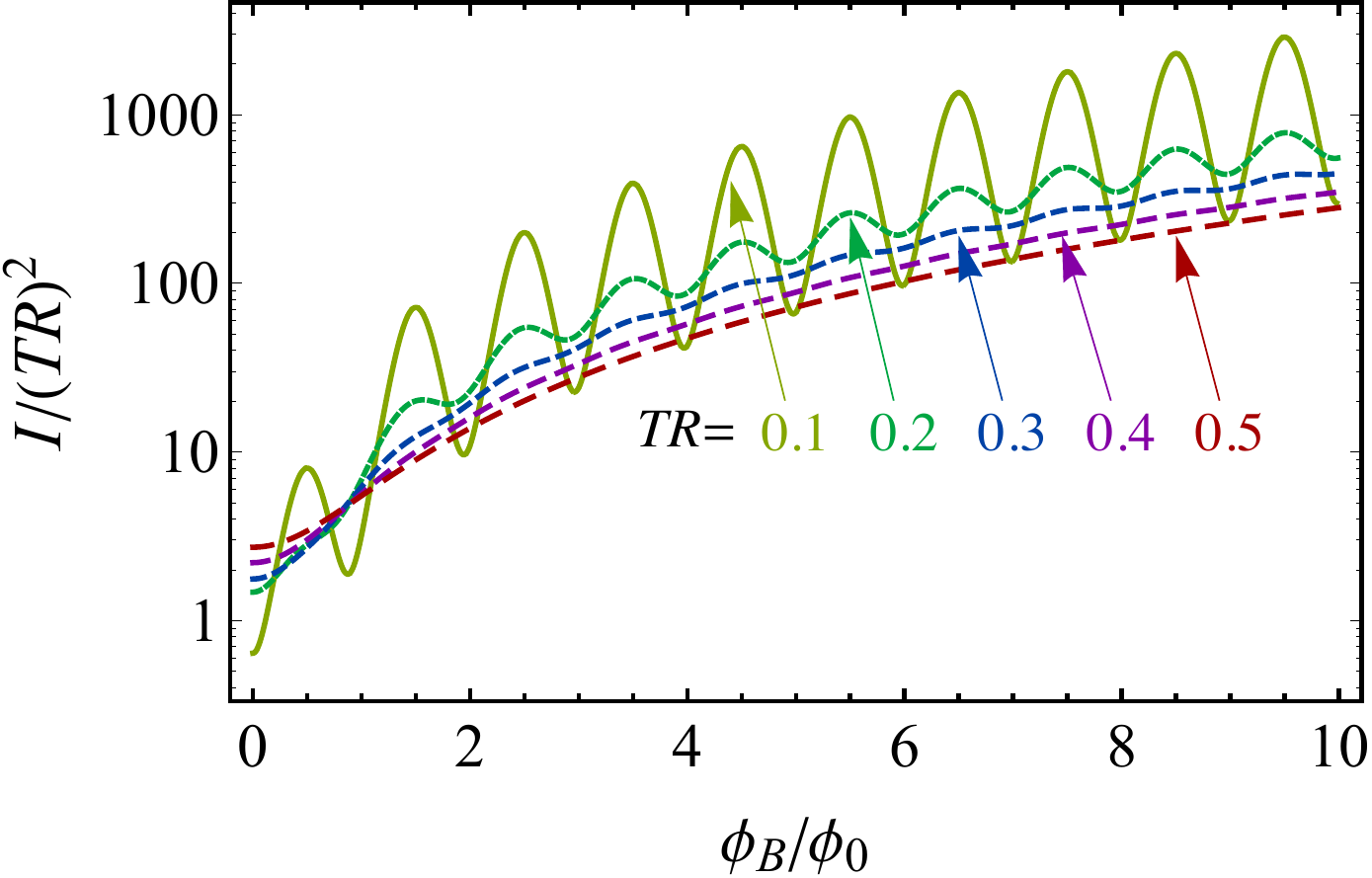}
\caption{Moment of inertia (divided by temperature squared) per unit height of cylinder vs. the flux $\phi_B$ of the background magnetic field at zero angular frequency $\Omega=0$.}
\label{fig:moment:inertia:B}
\end{figure}

In Fig.~\ref{fig:moment:inertia:B} we show the dependence of the moment of inertia (normalized by the temperature squared) at vanishing angular frequency $\Omega=0$ vs.  normalized magnetic flux~\eq{eq:phi:B:0}. We clearly see that that with increase of temperature the moment of inertia of the edge modes increases in agreement with zero-field behavior shown in Fig.~\ref{fig:moment:inertia:B0}. Similarly to the angular momentum, the moment of inertia experiences (nonperiodic) oscillations as a function of magnetic field. The local minima (maxima) approximately correspond to the integer (half-integer) values of the magnetic flux [calculated in units of the elementary flux~\eq{eq:phi:B:0}]. The oscillatory quantum behavior is well pronounced at low temperatures while at higher temperatures the dependence of the moment of inertia on the magnetic flux reduces to a monotonically increasing function. These features are also well visible in the plot~\eq{fig:moment:inertia:BL3d} which shows the moment of inertia $I$ vs. both magnetic flux $\phi_B$ and temperature $T$.

\begin{figure}[!thb]
\includegraphics[scale=0.5,clip=true]{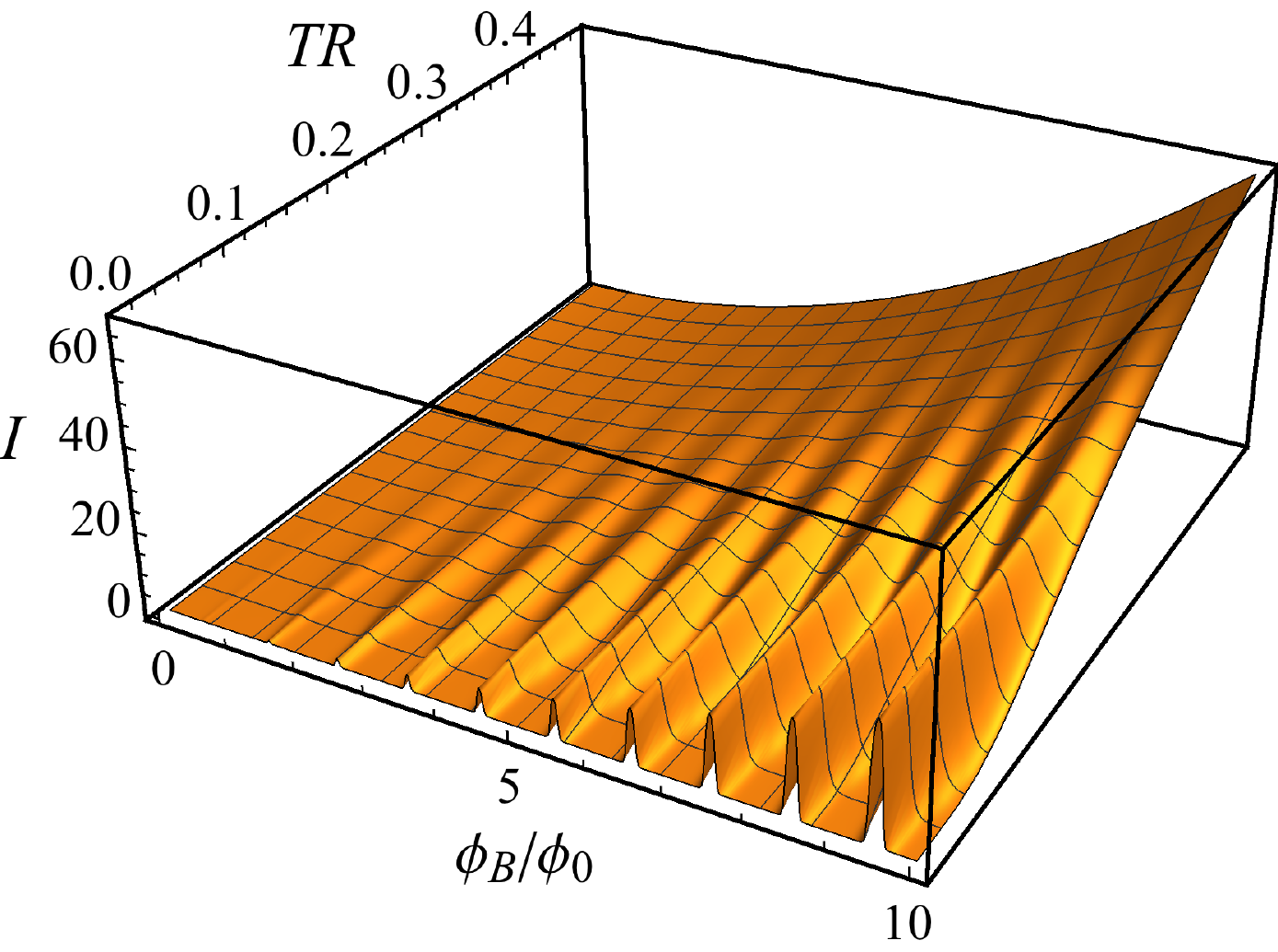}
\caption{Moment of inertia  per unit height of cylinder vs. magnetic flux $\phi_B$ and temperature $T$ at zero angular frequency $\Omega=0$.}
\label{fig:moment:inertia:BL3d}
\end{figure}

The fact that both the moment of inertia and the angular momentum are not periodic function of magnetic field is a natural consequence of non-equivalence of magnetic field and rotation in relativistic domain. Indeed, in many non-relativistic quantum-mechanical applications a (slow) rotation may be treated as a (weak) magnetic field. This fact is used, for example, in characterizing the spectrum of rotation optical lattices of cold atoms~\cite{ref:Zoller}. The equivalence is no more true in the case of a fast relativistic rotation: the effects of rotation and magnetic field in this case are very different~\cite{Chen:2015hfc,Chernodub:2016kxh}. In order to highlight the difference between rotation and magnetic field we mention that the ground state degeneracy is independent of the value of the angular frequency contrary to case of magnetic field~\cite{Chernodub:2016kxh}. Moreover, the phenomena of dimensional reduction, which govern many interesting effects in magnetic field background, does not exist in the case of rotation~\cite{Chernodub:2016kxh}.

\section{Conclusions}

We study a uniformly rotating relativistic system of free Dirac fermions in the background of a constant magnetic field directed along the axis of rotation. The system must be bounded in any plane perpendicular to the rotation axis in order to respect the relativistic causality according to requirement that the rotational velocity of particles does not exceed the speed of light. Therefore we enclose the system into an infinitely high cylinder of radius $R$ and restrict the angular frequency $\Omega$ of rotation to the subluminal domain: $\Omega R < 1$. At the surface of the cylinder we impose either the MIT boundary condition~\eq{eq:MIT:bc} or its chiral generalization~\eq{eq:chiral:theta} which is characterized by the chiral angle~$\Theta$. Both these conditions force the normal component of the fermionic current to vanish at cylinder's surface thus conserving the global fermionic number inside the rotating cylinder. 

In general, the spectrum of fermions in a finite geometry contains two types of solutions: bulk solutions concentrated in the interior of the system and the edge states which are localized at the boundary. The bulk states in cylindrical geometry were already discussed in the literature. In the absence of magnetic field the bulk spectrum of fermions was obtained in Ref.~\cite{Ambrus:2015lfr} where the cylinder with the MIT boundary conditions~\eq{eq:MIT:bc} was studied. The bulk spectrum with the chiral MIT boundary conditions~\eq{eq:chiral:theta} was found later in Ref.~\cite{Chernodub:2017ref}. In our paper we extend these results in various directions. 

Firstly, we find that the system possesses the edge modes at certain region of the parameter space. Secondly, we extend the results for the edge and bulk modes to the case of nonzero magnetic field  parallel to the axis of the cylinder (so that the magnetic flux is a constant quantity along the axis of the cylinder). Thirdly, we implement the uniform rotation of the whole system and investigate the interplay between rotation and magnetic field in thermodynamical properties of free fermions.\footnote{Uniformly rotating fermions in magnetic field were also studied in Ref.~\cite{Chen:2015hfc} in a transversally unrestricted geometry which does not possess the edge modes.} Fourthly, we highlight the role of the edge states that were neglected so far in the analysis of thermodynamics of rotating fermionic systems.

We found the following features of the system:

\begin{enumerate}

\item {\emph{The boundary condition is important for the edge states.}} The mass spectrum and the very existence of the edge modes depend on the values of the fermion mass $M$, magnetic field $B$ and the chiral $\Theta$ angle at the boundary. For example, there are no edge states at the chiral angle $\Theta=\pi/2$ at zero magnetic field.

\item {\emph{The lowest (ground-state) bulk modes transform into the edge states and vice-versa}} as the value of the fermion mass $M$ crosses, for each fixed value of the angular momentum~\eq{eq:mu:m}, a certain threshold mass. In the absence of magnetic field the threshold masses~\eq{eq:M:c} are given, for the MIT boundary conditions~\eq{eq:MIT:bc}, by $M_c = - n/R$ with $n = 1,2,\dots$. They differ from the threshold masses for the fermions with the chiral boundary conditions~\eq{eq:M:c:theta}. The threshold masses for the MIT boundary conditions are changed to Eq.\eq{eq:zero:M} in the case of nonzero magnetic field.

\item {\emph{The edge states are massive}} so that in the solid-state language the system may be associated with a non-topological insulator. 

\item {\emph{The masses of the edge states are finite for $B=0$.}} In the absence of magnetic field the spectrum is degenerate with respect to the sign flips of the angular momentum, $\mu_m \to - \mu_m$, see Fig.~\ref{fig:masses}. The masses of the bulk (edge) modes rise (fall) with increase of the absolute value of the fermion mass $M$. In the limit of a negative infinite fermionic mass, $M \to - \infty$, the bulk modes become infinitely heavy so that they decouple from the dynamics of the system and disappear. On the contrary, in this limit the masses of the edge modes remain finite~\eq{eq:M:edge:infty}. They are proportional to the mean curvature of the cylinder's surface, $1/R$.

\item {\emph{The masses of the edge states may vanish for $B \neq 0$.}} Nonzero magnetic field lifts out the $\mu_m \to - \mu_m$ degeneracy of the mass spectrum of both the bulk states and the edge states, see Fig.~\ref{fig:masses:magnetic}. For example, the edge states with $\sign{\mu_m eB}>0$ possess only nonvanishing masses while the masses of the edge states with $\sign{\mu_m eB}>0$ may become zero at certain values of momentum, shown in Fig.~\ref{fig:M:crit}. The masses of the bulk modes become infinitely massive in the limit $M \to -\infty$ while the masses of the edge states exhibit a periodic dependence on the magnetic flux, see Fig.~\ref{fig:edge:masses:infinite}, described by the simple formula~\eq{eq:M:edge:infty:B}.

\item {\emph{Moment of inertia oscillates with magnetic field.}} The presence of magnetic field affects drastically the rotational properties of the system. For example, in the domain of low temperatures in the limit of infinitely large negative fermion mass -- where the thermodynamics is given by the edge modes only -- the angular momentum (Fig.~\ref{fig:angular:momentum:B:Omega}) and, consequently, the moment of inertia (Fig.~\ref{fig:moment:inertia:B}) experience quasi-periodic (quantum) oscillations as functions of magnetic flux $\phi_B$. The local minima (maxima) of the moment of inertia correspond to the integer (half-integer) values of the magnetic flux $\phi_B$ in units of the elementary flux $\phi_0$, Eq.~\eq{eq:phi:B:0}. At high temperature the oscillations disappear, see Fig.~\ref{fig:moment:inertia:B}.
\end{enumerate}

\acknowledgments
The authors are grateful to Pavel Buividovich, Mark Goerbig and Mar\'ia Vozmediano for discussions. The work of S.~G. was supported by the Special Postdoctoral Researchers Program of RIKEN.

\end{document}